\documentclass[fleqn,usenatbib]{mnras}

\usepackage{newtxtext,newtxmath}
\usepackage[T1]{fontenc}

\DeclareRobustCommand{\VAN}[3]{#2}
\let\VANthebibliography\thebibliography
\def\thebibliography{\DeclareRobustCommand{\VAN}[3]{##3}\VANthebibliography}

\usepackage{graphicx}	% Including figure files
\usepackage{amsmath}	% Advanced maths commands
\usepackage{pifont}
\usepackage{multirow}
\usepackage{tabularx}

\newcommand{\tranf}{{\sc TransitFit}}
\newcommand{\plc}{{\sc PyLightcurve}}
\newcommand{\batman}{{\sc batman}}

\newcommand{\rjup}{R_\text{Jup}}
\newcommand{\mjup}{M_\text{Jup}}

\DeclareUnicodeCharacter{2212}{-}

\title[TransitFit]{TransitFit: combined multi-instrument exoplanet transit fitting for JWST, HST and ground-based transmission spectroscopy studies}

\author[J. J. C. Hayes et al.]{J.J.C. Hayes,$^{1}$
A.~Priyadarshi,$^{1}$\thanks{Corresponding author: Akshay.Priyadarshi@manchester.ac.uk}
E.~Kerins,$^{1}$\thanks{Corresponding author: Eamonn.Kerins@manchester.ac.uk}
S.~Awiphan,$^{2}$
I.~McDonald,$^{1,3}$
N.~A-thano,$^{4}$
J.S.~Morgan,$^{1}$\newauthor
A.~Humpage,$^{1}$
S.~Charles,$^{1}$
M.~Wright,$^{1}$
Y.C.~Joshi,$^{5}$
Ing-Guey~Jiang,$^{4}$
T.~Inyanya,$^{6,7}$
T.~Padjaroen,$^{6}$\newauthor
P.~Munsaket,$^{8}$
P.~Chuanraksasat,$^{2,9}$
S.~Komonjinda,$^{6}$
P.~Kittara,$^{10}$
V.S.~Dhillon,$^{11,12}$
T.R.~Marsh,$^{13}$\newauthor
D.E.~Reichart,$^{14}$ and
S.~Poshyachinda$^{2}$ (The SPEARNET Collaboration)
\\
$^{1}$Jodrell Bank Centre for Astrophysics, University of Manchester, Oxford Road, Manchester, M13 9PL, UK\\
$^{2}$National Astronomical Research Institute of Thailand, 260 Moo 4, Donkaew, Mae Rim, Chiang Mai, 50180, Thailand\\
$^{3}$Department of Physical Sciences, The Open University, Walton Hall, Milton Keynes, MK7 6AA, UK\\
$^{4}$Department of Physics and Institute of Astronomy, National Tsing-Hua University, Hsinchu 30013, Taiwan\\
$^{5}$Aryabhatta Research Institute of Observational Sciences (ARIES), Manora peak, Nainital - 263002, India\\
$^{6}$Department of Physics and Materials Science, Faculty of Science, Chiang Mai University, Chiang Mai, 50200, Thailand\\
$^{7}$Yupparaj Wittayalai School, Chiang Mai, 50200, Thailand\\
$^{8}$School of Physics, Institute of Science, Suranaree University of Technology, 111 University Ave., Suranaree, Nakhon Ratchasima 30000, Thailand\\
$^{9}$Department of Civil and Environmental Engineering, National University of Singapore, 1 Engineering Drive 2, 117576, Singapore\\
$^{10}$Department of Physics, Faculty of Science, Mahidol University, Bangkok 10400, Thailand\\
$^{11}$Department of Physics and Astronomy, University of Sheffield, Sheffield S3 7RH, UK\\
$^{12}$Instituto de Astrof\'{i}sica de Canarias, La Laguna, Tenerife E-38205, Spain\\
$^{13}$Department of Physics, University of Warwick, Coventry CV4 7AL, UK\\
$^{14}$Department of Physics and Astronomy, University of North Carolina at Chapel Hill, Chapel Hill, NC 27599, USA\\
}

\date{Accepted XXX. Received YYY; in original form ZZZ}

\pubyear{2023}

\begin{document}
\label{firstpage}
\pagerange{\pageref{firstpage}--\pageref{lastpage}}
\maketitle

\begin{abstract}

We present \tranf{}\footnotemark, a package designed to fit exoplanetary transit light-curves. \tranf{} offers multi-epoch, multi-wavelength fitting of multi-telescope transit data. \tranf{} allows per-telescope detrending to be performed simultaneously with transit parameter fitting, including custom detrending. Host limb darkening can be fitted using prior conditioning from stellar atmosphere models.  We demonstrate \tranf{} in a number of contexts. We model multi-telescope broadband optical data from the ground-based SPEARNET survey of the low-density hot-Neptune WASP-127~b and compare results to a previously published higher spectral resolution GTC/OSIRIS transmission spectrum.
Using \tranf{}, we fit 26 transit epochs by \emph{TESS} to recover improved ephemeris of the hot-Jupiter WASP-91~b and a transit depth determined to a precision of 111~ppm. We use \tranf{} to conduct an investigation into the contested presence of TTV signatures in WASP-126~b using 180 transits observed by \emph{TESS}, concluding that there is no statistically significant evidence for such signatures from observations spanning    27 \emph{TESS} sectors. We fit \emph{HST} observations of WASP-43~b, demonstrating how \tranf{} can use custom detrending algorithms to remove complex baseline systematics. Lastly, we present a transmission spectrum of the atmosphere of WASP-96~b constructed from \emph{simultaneous} fitting of \emph{JWST} NIRISS Early Release Observations and archive \emph{HST} WFC3 transit data. The transmission spectrum shows generally good correspondence between spectral features present in both datasets, despite very different detrending requirements.
\end{abstract}


%%%%%%%%%%%%%%%%%%%%%%%%%%%%%%%%%%%%%%%%%%%%%%%%%%

%%%%%%%%%%%%%%%%% BODY OF PAPER %%%%%%%%%%%%%%%%%%

\footnotetext{Available at \url{https://github.com/SPEARNET/TransitFit}, with documentation available at \url{https://transitfit.readthedocs.io/en/latest/}.}

\section{Introduction}
\label{sec:intro}
Over the last decade, the study of exoplanetary atmospheres though transmission spectroscopy studies has been a growing and maturing field, seeing a significant increase in the number of surveys targeting transiting exoplanets. Dedicated space-based transit surveys, such as the \textit{Kepler Space Telescope} \citep{kepler}, and the more recent \textit{Transiting Exoplanet Survey Satellite} \citep[\textit{TESS},][]{TESS_paper} together with ground-based surveys such as the \textit{Next Generation Transit Survey} \citep[NGTS, ][]{NGTS}, have provided an ever-growing list of targets for study. These surveys have contributed significantly to the growing number of confirmed exoplanets: at the time of writing, there are over $5500$ confirmed exoplanets, with over $3900$ of these exhibiting observable transits\footnote{The Extrasolar Planet Encyclopedia: \url{http://exoplanet.eu/}}. Early surveys, such as the \textit{Wide Angle Search for Planets} \citep[WASP, ][]{WASP_survey}, were ground-based and therefore most of the exoplanets discovered in the early days of the field were bright enough to allow for ground-based follow-up studies.

Many of the planets discovered by recent surveys such as TESS can be observed with both ground- and space-based telescopes, and as such, the field of transmission spectroscopy is leaving the ``target-starved'' era and entering an ``asset-starved'' era. The limiting factor to exoplanetary studies is now no longer the availability of viable targets, but instead the availability of ground-based facilities to conduct follow up. To adapt to this change, new tools and techniques need to be developed to best utilise available resources.

Currently a variety of atomic and molecular species have been identified through analysis of transmission spectra, including potassium \citep[i.e.][]{potassium2011, potassium2015}, sodium \citep{charbonneau2002detection, redfield2008sodium_detection}, water \citep{tinetti2007water, grillmair2008strong, swain2008methane, konopacky2013CO_H2O_detection, Birkby_water_2013}, titanium oxide \citep{sedaghati2017TiOdetection, NugruhoTiO}, carbon monoxide \citep{Snellen_CO_2010, Brogi_CO_2012}, HCN \citep{tsiaras2016detection, Hawker_HCN}, methane \citep{swain2008methane, guilluy_CH4}, helium \citep{Allart_He}, vanadium oxide \citep{evans2016water_TiO/VO_detection}, iron \citep{Hoeijmakers_Ti_Fe}, and carbon dioxide \citep{2023Natur.614..649J}.

The identification of these species relies on two stages of retrieval. First, a spectrum must be acquired from measurements of the radius of a planet at different observation wavelengths, and then an atmospheric model can be used to obtain atmospheric parameters for the planet. Obtaining accurate planet-host radius ratios from light curves is a significant challenge as many factors affect the shape of a transit light curve, including atmosphere and orbital parameters of the exoplanet, the behaviour of the host, and instrumental and terrestrial factors.

There are currently a few publicly available codes designed for fitting light curves of transiting exoplanets. \plc{} \citep{pylightcurve} is a complete forward model and retrieval package, which uses an MCMC routine to fit transit light curves. \plc{} can also simultaneously remove trends from data using a $2$nd-order polynomial and offers a variety of limb-darkening laws. {\sc exoplanet} \citep{exoplanet_python_package} is a toolkit for modelling transit and radial velocity observations of exoplanets, and is built with multi-planet systems in mind. Similarly to \plc{}, it uses MCMC methods to fit transit curves, and has a variety of limb-darkening laws, though it does not offer detrending functionality. {\sc ExoFastv2} \citep{2019arXiv190709480E} is an IDL package that can perform simultaneous fitting of transit, radial velocity and astrometric observations. It is highly optimized and offers a host of user options, including simultaneous detrending, using an MCMC optimizer.

Approaches to fitting limb-darkening coefficients (LDCs) vary, with some researchers fixing values before retrieval and some instead fitting them as free parameters as part of retrieval. \citet{LDC_fitting_in_exoplanets} discuss this and conclude that it is best to freely-fit LDCs as fixing them can lead to biases of up to $3$~per~cent in measurements of $R_p/R_\star$, which can have significant effects on retrieved spectra. \citet{WASP-127b_NaK&Li} demonstrated that by using information about the host star, namely temperature, mass (or surface gravity), and metallicity, it is possible to improve the fitting of LDCs and consequently the fitting of transit light curves in general. Whilst fitting LDCs as free parameters is common practise, it is clear that LDCs depend on observation wavelength and on the properties of the host star and therefore it is worth trying to develop tools that can exploit this additional information.

In this paper, we present \tranf{}, an open-source Python 3 package for robust multi-wavelength, multi-epoch fitting of transiting exoplanet light curves obtained from one or more telescopes. \tranf{} has been developed in response to the fast-growing numbers of available transmission spectroscopy targets, as part of \textit{the Spectroscopy and Photometry of Exoplanetary Atmospheres Research Network} (SPEARNET), a survey that is employing automated transmission spectroscopy target selection for follow-up by a globally dispersed and heterogeneous telescope network \citep{morgan2019metric}. In Section \ref{sec:implementation}, we discuss the implementation of \tranf{}, including the approach to limb-darkening, simultaneous detrending, and its handling of large multi-epoch, multi-wavelength datasets. In Section \ref{sec:application} we demonstrate the application of \tranf{} to five different situations. We illustrate the use of \tranf{} on multi-telescope, multi wavelength, multi-epoch observations of WASP-127~b obtained by SPEARNET. We then look at two examples of the use of multi-epoch \emph{TESS} data, producing improved ephemeris for WASP-91~b and conducting a sensitive investigation into the presence of contested TTV signatures from WASP-126~b. Finally, we demonstrate how \tranf{} can be used to fit complex systematics through analysis of \emph{HST} observations of WASP-43~b, and the capability of \tranf{} to fit \emph{JWST} NIRISS and \emph{HST} WFC3 observations of WASP-96~b. We offer our conclusions in Section \ref{sec:conclusions}.

\section{Implementation of \tranf{}}
\label{sec:implementation}
\tranf{} is an open-source, pure Python 3.x package designed specifically with transmission spectroscopy studies in mind and uses transit observations at different wavelengths and epochs from different telescopes simultaneously to fit transit parameters using nested sampling retrieval. This approach allows lightcurves to be fitted by exploiting coupled information across wavelength and epoch. \tranf{} offers a wavelength-coupled approach to LDC fitting, which we discuss in depth in Section \ref{sec:limb_darkening}. We also discuss how \tranf{} can detrend and normalise light curves simultaneously with fitting other parameters in Section \ref{sec:detrending}. Different light curves being fitted simultaneously may benefit from using different detrending functions, especially if obtained from different telescopes. \tranf{} can also deal with multi-epoch observations from systems which exhibit transit timing variations (TTVs), although, as highlighted in Section \ref{sec:ttvs}, there are some current limitations to fitting these systems. 

The forward model used to calculate transit light curves in \tranf{} is the \batman{} Python package \citep{batman}. We have chosen \batman{} as it is well-established within the community, and the open-source nature of it means that we can easily incorporate it into an open-source retrieval code. Since the choice of forward model is what limits the parameters which can be fitted, using \batman{} means that the physical parameters retrievable by \tranf{} are: orbital period, $P$; the time of inferior conjunction, $t_0$; the planet-star radius ratio, $R_p/R_\star$; the ratio of semi-major axis of the planet orbit to the radius of the star, $a/R_\star$; the orbital inclination, $i$; the orbital eccentricity, $e$; the longitude of periastron for the planet's orbit, $\omega$; and a set of limb-darkening coefficients, written as a matrix $\mathbf{U}$, where $U_{i,j}$ is the $j$th LDC for filter $i$. Along with these, \tranf{} can calculate detrending parameters and normalisation constants for individual light curves. 

Retrieval is conducted using nested sampling routines \citep{nested_sampling_2004, skilling2006} implemented with the {\sc dynesty} Python package \citep{dynesty}. 
During each iteration of fitting, we use \batman{} Python package to calculate model light curve from the sampled parameters, while simultaneously detrending and normalising the raw lightcurves using the parameters sampled in the same iteration. 
The likelihood function for the fitting includes an additional error-rescaling term ($f_e$) as described in the online documentation for \textsc{EMCEE}\footnote{\url{https://emcee.readthedocs.io/en/stable/tutorials/line/}} \citep{2013PASP..125..306F}. This allows for cases where the flux errors might be underestimated.
The retrieval returns the highest-likelihood parameter values as the best fit parameters. We define upper limit on the best fit value as the 68.27 quantile of the weighted samples beyond the best fit value, and the lower limit as 31.73 quantile of weighted samples below the best fit.

General use of \tranf{} involves 3 configuration files, one directing \tranf{} to the data to be fitted, one defining the priors, and one defining the observation filter profiles, and a single wrapper function, \texttt{run\_retrieval}, which handles under-the-hood code interfacing. In order to be able to denote individual observations and their relationships, each light curve passed to \tranf{} is identified by three indices: telescope, filter, and epoch. Observations which come from the same telescope will share a telescope index, and so on. By taking this three-index approach, \tranf{} is able to easily fit shared parameters, such as $R_p$ values for observations made in the same filter. In addition to the telescope, filter, and epoch indices, each observation is given an index referring to a specific detrending model (see Section \ref{sec:detrending}). The models that each detrending index refers to is set by the user, and allows the use of more than one detrending model across a set of observations. 

\subsection{Limb-darkening}
\label{sec:limb_darkening}
\tranf{} uses the Limb Darkening Toolkit (LDTk) \citep{Parviainen2015} to calculate the likelihood of sets of limb-darkening coefficients given planet host characteristics and the wavelength of observations. This allows limb darkening priors to be constructed that tension (though not fix) the wavelength-dependent behaviour of limb darkening in accordance with the {\sc Phoenix} stellar atmosphere models \citep{Husser2013}. As well as this ``coupled'' approach, \tranf{} can compute limb-darkening coefficients independently for each filter, corresponding to the more traditional ``uncoupled'' approach that allows for easy comparison with other models. As a third option, \tranf{} can use a ``fixed'' mode where LDC are fit in only one waveband and LDTk is used to compute LDC values for other observed wavebands. This latter mode is offered for use in situations where coupling is desired, but a large number of filters leads to an unreasonable number of parameters to be fitted. However, such a fixed LDC approach can lead to biased results \citep{LDC_fitting_in_exoplanets}. We therefore advocate using \tranf{} in coupled mode whenever feasible.

\subsubsection{Limb-darkening models}
\label{sec:LD_models}
Stellar intensity varies between the centre and the edge of the disk, which often results in the base of a transit curve being rounder than if the stellar disk were a uniform brightness. Typically, these variations in intensity are described by analytical functions $I_\lambda\left(\mu\right)$, where $\mu$ is the cosine of the angle between the line of sight and the emergent intensity. $\mu$ can also be expressed as $\mu = \sqrt{1-r^2}$ where $r$ is the unit-normalised radial coordinate on the stellar disk, and as such, all limb-darkening models must be valid for $0 \le \mu < 1$.

There are a wide variety of limb-darkening models \citep{Claret2000}, and several have been implemented in \tranf{}. These are the linear law \citep{linear_LD_law}
\begin{equation}
    \frac{I\left(\mu\right)}{I\left(1\right)} = 1 - u_{0,l} \left(1 - \mu\right),
    \label{eq:linearLD}
\end{equation}
the quadratic law \citep{KopalQuadraticLimbDarkening},
\begin{equation}
    \frac{I\left(\mu\right)}{I\left(1\right)} = 1 - u_{0,q} \left(1 - \mu\right) - u_{1,q} \left(1-\mu\right)^2,
    \label{eq:quadraticLD}
\end{equation}
the square-root law \citep{sqrtLD},
\begin{equation}
    \frac{I\left(\mu\right)}{I\left(1\right)} = 1 - u_{0,\textrm{sqrt}} \left(1 - \mu\right) - u_{1,\textrm{sqrt}} \left(1-\sqrt{\mu}\right),
    \label{eq:sqrtLD}
\end{equation}
the power-2 law \citep{power2law}
\begin{equation}
    \frac{I\left(\mu\right)}{I\left(1\right)} = 1 - u_{0,\textrm{p2}}\left(1 - \mu^{u_{1,\textrm{p2}}}\right),
    \label{eq:power2LD}
\end{equation}
and the non-linear law \citep{Claret2000}
\begin{equation}
    \begin{split}
        \frac{I\left(\mu\right)}{I\left(1\right)} = 1 & - u_{0,\textrm{nl}} \left(1 - \mu^{1/2}\right) - u_{1,\textrm{nl}} \left(1-\mu\right) \\
        &- u_{2,\textrm{nl}} \left(1-\mu^{3/2}\right) - u_{3,\textrm{nl}} \left(1-\mu^{2}\right).
    \end{split}
    \label{eq:nonlinearLD}
\end{equation}
Each of $u_0$, $u_1$, $u_2$, and $u_3$ are the LDCs, which must be fitted simultaneously with other transit parameters during retrieval. The LDCs are, however, dependent upon wavelength, and consequently must be fitted for each filter used in an observation or set of observations.

\subsubsection{Constraining limb-darkening coefficients}
\label{sec:constraining_LDCs}
The most basic approach for fitting limb-darkening coefficients is to sample them independently and find the best fit values, but this can allow unphysical values to be trialled with no penalty. This is an issue which was addressed by \citet{KippingLimbDarkening}, who imposed two conditions on limb-darkening profiles to ensure that they are physically allowed:
\begin{enumerate}
    \item \label{condition1}The intensity profile must be always positive, or
    \begin{equation}
        \frac{I\left(\mu\right)}{I\left(1\right)} >0~~\forall~~0 \le \mu < 1
    \end{equation}
    
    \item \label{condition2} The intensity profile must be monotonically decreasing from the centre to the edge of the stellar disk, meaning that 
    \begin{equation}
    \frac{\partial I\left(\mu\right)}{\partial \mu}  > 0 ~~\forall~~0 \le \mu < 1.
\end{equation}
\end{enumerate}
\citet{KippingLimbDarkening} showed that by applying these conditions, it is possible to place constraints on allowed values of the LDCs for the two-parameter quadratic and square root laws. To improve the efficiency of sampling within this restricted space, rather than sampling the LDCs $\left\{u_0, u_1\right\}$, \citet{KippingLimbDarkening} instead reparameterises the laws in terms of the coefficients $\left\{p\left(u_0, u_1\right), q\left(u_0, u_1\right)\right\} \in \left[0, 1\right]$. This reparameterisation ensures a sampling efficiency of $100$~per~cent (i.e. all samples are within the physically allowed region), without the need for checking that sampled values of $u_0$ and $u_1$ follow the imposed constraints. An alternate way of parameterisation for the power-2 law coefficients has been discussed in \citet{Short_2019}.

We have implemented the parameterisation following \citet{KippingLimbDarkening} in \tranf{}, and have extended the method to the power-2 law in Equation \ref{eq:power2LD}. In the limit $\mu \to 0$, Condition \ref{condition1} yields the constraint 
\begin{equation}
   u_{0,\textrm{p2}}< 1.
\end{equation}
Condition \ref{condition2} implies that 
\begin{equation}
    \frac{\partial I\left(\mu\right)}{\partial \mu} = u_{0,\textrm{p2}}u_{1,\textrm{p2}}\mu^{u_{1,\textrm{p2}}-1} > 0 ~~\forall~~0 \le \mu < 1.
\end{equation}
This does not give us anything overtly useful in the limit $\mu \to 0$ due to the cross terms, however, as $\mu \to 1$, we see that
\begin{equation}
    u_{0,\textrm{p2}}u_{1,\textrm{p2}} > 0.
\end{equation}
This places the constraint that the power-2 LDCs must have the same sign, and can be viewed as a `quadrant limiting' constraint. Mathematically, there is no lower bound on $u_{0,\textrm{p2}}$, and there are no bounds at all on $u_{1,\textrm{p2}}$ but, from a computational perspective, we must place limits on them in order to be able to sample values. Therefore, we can say that $u_{0,\textrm{p2}} \in \left[u_{0,\textrm{p2}}^{min}, 1\right]$ and $u_{1,\textrm{p2}} \in \left[u_{1,\textrm{p2}}^{min}, u_{1,\textrm{p2}}^{max}\right]$. We implement this by fitting $\left\{p_{p2}, q_{p2}\right\} \in \left[0, 1\right]$ and using the conversions 
\begin{equation}
    u_{0,\textrm{p2}} = p_{p2}\left(1 - u_{0,\textrm{p2}}^{min}\right) + u_{0,\textrm{p2}}^{min}
\end{equation}
and 
\begin{equation}
    u_{1,\textrm{p2}} = 
    \begin{cases}
    u_{1,\textrm{p2}}^{min}\left(1 - q_{p2}\right), &\text{for } u_{0,\textrm{p2}} < 0\\
    u_{1,\textrm{p2}}^{max}q_{p2}, &\text{for } u_{0,\textrm{p2}} \ge 0
    \end{cases}.
\end{equation}
Provided that $u_{1,\textrm{p2}}^{min} = -u_{1,\textrm{p2}}^{max}$, it can be shown that this method uniformly samples in all the allowed regions in the $\left\{u_{0,\textrm{p2}}, u_{1,\textrm{p2}}\right\}$ plane, with 100~per~cent efficiency. 

It is trivial to also apply this method to the linear law, which places the constraint
\begin{equation}
    0 < u_{0,\textrm{l}} < 1,
\end{equation}
and this has been implemented in \tranf{}. However, this method has yet to be successfully applied to the non-linear law. The current best attempt is by \cite{Kipping_2015}, where the methodology is extended to the three-term law of \citet{Sing_3param_LD}, which drops the $\mu^{1/2}$ term from the non-linear law in Equation \ref{eq:nonlinearLD}. Consequently, \tranf{} does not use the Kipping parameterisation to limit sampling of non-linear LDCs to a physically-allowed region of parameter space. 

\subsubsection{Coupling limb-darkening coefficients across wavelengths}
\label{sec:coupling_LDCs}
Multiple codes use the Kipping parameterisation to constrain the LDC values to those which are physically allowed, but it is possible to improve the quality of LDC fitting further. All of the currently-available transit-fitting codes fit LDCs for each filter independently. This means that for each filter a transit is observed at, the best-fit LDCs for each filter may not be physically consistent with each other. \cite{Parviainen2015} developed the Limb Darkening Toolkit (LDTk) to allow researchers to address this problem, but we have been unable to find a publicly available transit fitting code that makes use of LDTk.

LDTk uses the library of PHOENIX stellar atmospheres and synthetic spectra \citep{Husser2013} to estimate the likelihood of a set of stellar LDCs for a given set of observation filters. Using this, we have given \tranf{} the functionality to couple LDCs across multiple filters, which can then be fitted simultaneously. This allows for the refinement of the limb-darkening physics included in transmission spectroscopy studies by ensuring that the retrieved LDC values are statistically tensioned across filters in a manner consistent with stellar atmosphere models. The filter profiles used by \tranf{} can be either uniform, box filter profiles, which may be suitable to represent individual spectroscopic channels, or user-supplied filter profiles for broadband photometric studies. In the case where a specific filter profile cannot be obtained, we recommend using the equivalent width of a filter. \tranf{} is distributed with filter profiles for the Johnson-Cousins UVBRI set, and the SLOAN-SDSS \textit{u'g'r'i'z'} set, as well as profiles for \textit{Kepler} and \textit{TESS}. All of these were obtained from the SVO Filter Profile Service\footnote{\url{http://svo2.cab.inta-csic.es/theory/fps/}} \citep{filter_profile_service1, filter_profile_service2}.

Since there may be edge cases where this coupling is not desired, for instance where host information is unavailable, or computational limitations in the case of fitting a very large numbers of observation filters, such as spectroscopic observations, \tranf{} offers three modes of LDC fitting:
\begin{description}
    \item[\textit{Independent}:] This is the traditional approach of fitting LDCs for each filter separately. \tranf{} still uses the Kipping parameterisations laid out in Section \ref{sec:constraining_LDCs}, but LDTk is not used to couple LDCs across filters.
    \item[\textit{Coupled}:] Using the Kipping parameterisations, each LDC is fitted as a free parameter, with LDTk being used to estimate the likelihood of sets of LDCs, using information on the host star and the observation filters. \tranf{} also provides the functionality to use the uncertainty multiplier from LDTk.
    \item[\textit{Single filter}:] When fitting with multiple wavebands, the number of parameters required to be fitted can increase dramatically when using the coupled mode. Consequently, we have provided a method of only freely fitting the LDCs of one filter, and using LDTk to extrapolate LDC values for the remaining filters. For the $i$-th coefficient of a filter $f$, $c_{i, f}$, this extrapolation is calculated by
    \begin{equation}
        c_{i, f} = u_i \times \frac{\langle c_{i, f}\rangle}{\langle u_{i}\rangle}
        \label{eq:LDC_extrapolation}
    \end{equation}
    where $u_i$ is the sampled value of the $i$-th LDC in the actively fitted filter, and $\langle c_{i, f}\rangle$ and $\langle u_{i}\rangle$ are the maximum likelihood values initially suggested by LDTk.
\end{description}

\subsection{Detrending and normalisation of light curves}
\label{sec:detrending}
Transit light curves are sensitive to a variety of factors which stop the out-of-transit baseline being flat. These can range from host variations through to internal reflections within the telescope. These trends must be removed in order to obtain accurate parameters from observations. Additionally, many transit models, including \batman{}, normalise the out-of-transit baseline to a flux of 1. In some cases, detrending and normalisation is conducted before any further analysis, but ideally, detrending and normalisation coefficients should be fitted simultaneously with other model parameters to ensure that light curve features of interest are not inadvertently removed. 

In order to ensure that detrending does not bias our measurement of the transit depth we impose a constraint on the detrending function such that it conserves the relative flux at the epoch of mid-transit, $t_0$, (i.e we conserve the transit depth).
We refer to detrending functions which meet this criterion as ``depth-conserving.'' The enforcement of depth conservation provides an important constraint for normalising the detrending function. Similar relative photometry conservation is also at the heart of other differential photometry methods, such as difference imaging \citep{alard}. Note that since $t_0$ is itself a parameter that is being fitted for, the detrending correction and the $t_0$ determination are correlated and so are determined by \tranf{} simultaneously.

\tranf{} offers functionality to simultaneously detrend and normalise light curves during retrieval. Built into the package are depth-conserving, $n$-th order detrending functions, and the user can supply their own custom functions if more complicated detrending is required. 

The $n$-th order functions are calculated by writing the detrended flux values of a time series $\mathbf{t}$, $\mathbf{D}\left(\mathbf{t}\right)$ as 
\begin{equation}
    \mathbf{D}\left(\mathbf{t}\right) = \mathbf{F}\left(\mathbf{t}\right) - \mathbf{d}\left(\mathbf{t}\right)
\end{equation}
where $\mathbf{F}\left(\mathbf{t}\right)$ are the raw flux values and $\mathbf{d}\left(\mathbf{t}\right)$ is some detrending function. We place the constraint of depth conservation upon the detrended light curves such that 

\begin{equation} D\left(t_{0}\right)=F\left(t_{0}\right),\end{equation}
which gives us the constraint that 
\begin{equation}d\left(t_{0}\right)=0.\label{eq:sum_detrend_0} \end{equation}
By applying this conservation of transit depth at $\mathbf{t_0}$, we place constraints on the $0$-th component (intercept) of the detrending function. In the case of a linear detrending function, where 
\begin{equation}
    \mathbf{d}\left(\mathbf{t}\right) = a \mathbf{t} + b, 
    \label{eq:linear_detrending}
\end{equation}
applying Equation \ref{eq:sum_detrend_0} yields
\begin{equation}a t_{0}+b=0,\end{equation}
from which we can cast the $0$-th order term $b$ in terms of the other parameters to give 

\begin{equation}b=-a t_{0},\end{equation}
which can be substituted into Equation \ref{eq:linear_detrending}, resulting in 

\begin{equation}
    \mathbf{d}\left(\mathbf{t}\right) = a \left(\mathbf{t} - t_0\right).
\end{equation}
This can be generalised to $n$-th order (for $n > 0$) detrending functions given by 
\begin{equation}
    d\left(t_i\right) = \sum^n_{j=0} a_j t_i^j
    \label{eq:nth_order_detrending}
\end{equation}
as 

\begin{equation}
    d\left(t_i\right) = \sum^n_{j=1} \left[a_j \left(t_i - t_0\right)^j\right],
\end{equation}
where $a_j$ are the detrending coefficients and the exponent of the time series is bit-wise. 

This method allows us to also fit a normalisation constant without falling foul of degeneracy between the scaling due to the normalisation constant and the shift that a freely-fitted $0$-th order detrending term introduces. $0$-th order detrending cannot be applied due to this degeneracy, and we assume that these light curves are detrended, but not necessarily normalised. In the case of a user-defined detrending function, we strongly recommend following the above depth-conservation procedure in order to avoid the risk of degenerate solutions.

\subsection{Dealing with systems exhibiting TTVs}
\label{sec:ttvs}
The basic implementation of \tranf{} assumes that there are no transit timing variations (TTVs) within multi-epoch observations and fits one value of $t_0$, assumed to be consistent across all epochs. In the event that a system does exhibit TTVs, this method will fail to produce an accurate result. Consequently, in these cases, \tranf{} takes a slightly different approach.
\begin{enumerate}
    \item First, we consider each filter separately. We run retrieval on all the curves in this filter, using all the data to fit $R_p$ and limb-darkening coefficients. However, we fit a separate $t_0$ for each observation epoch within the filter, and cannot directly fit a period, $P$, in this mode, which must instead be provided. 
    \item Using the results from these single-filter retrievals, we detrend and normalise each light curve and then use the retrieved $t_0$ values to produce a phase-folded light curve for each filter. The observation times $\textbf{t}$ for each light curve are folded to give $\textbf{t}'$, where $t_0 - \frac{P}{2} < t' \le t_0 + \frac{P}{2}$, using 
    \begin{equation}
        \textbf{t}' = \textbf{t} - P\times \left\lfloor\frac{\textbf{t} - \left(t_0 + P/2\right)}{P}\right\rfloor - C
        \label{eq:folding_lightcurves}
    \end{equation}
    where
    \begin{equation}
        C = t_0 - P\times \left\lfloor\frac{t_0 - \left(t_0 + P/2\right)}{P}\right\rfloor - t_{0,\text{base}}
        \label{eq:ttv_folding_term}
    \end{equation}
    accounts for the offset caused by the different $t_0$ values for each epoch. By choosing a value for $t_{0,\text{base}}$ this term ensures that all the light curves are centred on $t_{0,\text{base}}$.
    
    \item Fit the folded light curves using the standard \tranf{} approach, coupling LDCs where required. 
\end{enumerate}
As stated above, when allowing for the presence of TTVs, \tranf{} cannot fit for $P$, which must be provided. For consistency, we recommend first running \tranf{} on data assuming that no TTVs are present, in order to obtain an appropriate value for $P$. \tranf{} cannot automatically detect TTVs in data, and must be instructed explicitly to allow for them. In the case where TTVs are present but \tranf{} is not allowing for them, the retrieved results will be incorrect. Additionally, \tranf{} does not solve the system dynamics associated with any present TTVs, as the purpose of \tranf{} is to produce robust transit fitting for the purposes of transmission spectroscopy studies.

\subsection{Batched retrieval: fitting a large number of parameters}
\label{sec:batched_retrieval}
As with any retrieval algorithm, increasing the dimensionality of the parameter space leads to instability in the nested sampling routines and can lead to inaccurate results. Since \tranf{} is anticipated to be used in transmission spectroscopy studies, where many tens, or even hundreds of light curves may need to be fitted, we have provided a solution to this in the form of ``batched'' retrieval.

In this mode, the user can specify the maximum number of parameters for \tranf{} to fit at one time. The light curves are then grouped by observation filter and split into multi-filter batches, where the number of parameters being fitted in each batch is less than the user-set limit. The batches are calculated to try and ensure that filters are present in multiple batches, which results in coupling between them. The exception to this is in the case that one filter has a high enough number of observations in it that the number of parameters required exceeds the user-set limit. In this case, this filter is fitted independently and does not benefit from any coupling. In these cases, we recommend using the ``folded'' mode.  After retrieval has been run on all the batches, a set of summary results are produced by calculating a weighted mean of all parameters. 

\subsection{Folded retrieval: Producing folded light curves}
With the launch of large surveys such as \textit{TESS}, many exoplanets have multiple-epoch observations in a single filter. \tranf{} can make use of these through a two-step retrieval process. In the first step, \tranf{} runs a retrieval on each filter independently, and uses the results to produce a phase-folded light curve for each filter. In the second step, \tranf{} runs a standard multi-wavelength retrieval using the batched algorithm above. This mode of retrieval allows for the production of high-quality folded light curves from non-detrended data, as well as providing a method where observations from long-term, single-waveband surveys such as \textit{TESS} can be easily combined with single-epoch observations at multiple wavelengths, such as from ground-based spectrographic follow-up.

\section{Applications of \tranf{}}
\label{sec:application}
\tranf{} was initially designed for use in spectroscopy studies, but also be applied to temporal studies, either in updating ephemeris of planets, or in studying systems for TTVs, which can be indicative of other planets in a system. 

In this section, we demonstrate the application of \tranf{} in four different scenarios, illustrating the impact of using LDTk to inform LDC fitting. First, we will discuss the fitting of multi-wavelength, ground based photometric observations of the low-density hot Neptune WASP-127~b \citep{WASP-127b_discovery}, using previously unpublished data acquired from the SPEARNET network of telescopes. We then move to applying \tranf{} to \textit{TESS} observations of the warm Jupiter WASP-91~b \citep{WASP-91b_discovery} and provide updated ephemeris and orbital parameters for the system. Third, we analyse \textit{TESS} observations of the hot Jupiter WASP-126~b \citep{WASP-126b_discovery}, a system which contentiously exhibits TTVs \citep{WASP-126b_TTVs, WASP-126c_no-evidence}, and use \tranf{} to show that there is no statistically significant evidence of TTVs within 180 transits observed by \textit{TESS}. Fourth, we analyse a single spectroscopic channel of the \emph{HST} observation of WASP-43~b made by \citet{WASP-43b_kreidberg_HST} to demonstrate the capability of \tranf{} to handle complex systematics through custom detrending functions. Finally, we fit the \emph{JWST} and \emph{HST} observations of WASP-96~b simultaneously to generate a transmission spectrum and show the capability of \tranf{} to work with \emph{JWST} observations. For all five systems, we assume circular orbits and use \tranf{} to fit the global parameters of $P$, $t_0$, $a/R_\star$, and $i$, as well as the filter-specific $R_p/R_\star$ and LDC values. We use an uncertainty multiplier of 10 for the LDC values, which is based on comparison of results from different grids in {\sc ExoCTK} \citep{matthew_bourque_2021_4556063} and {\sc ExoTiC-LD} \citep{david_grant_2022_7437681}.

\subsection{Broadband photometric observations of WASP-127~b}
\label{sec:WASP-127b}
\begin{figure*}
    \centering
    \includegraphics[width=\textwidth]{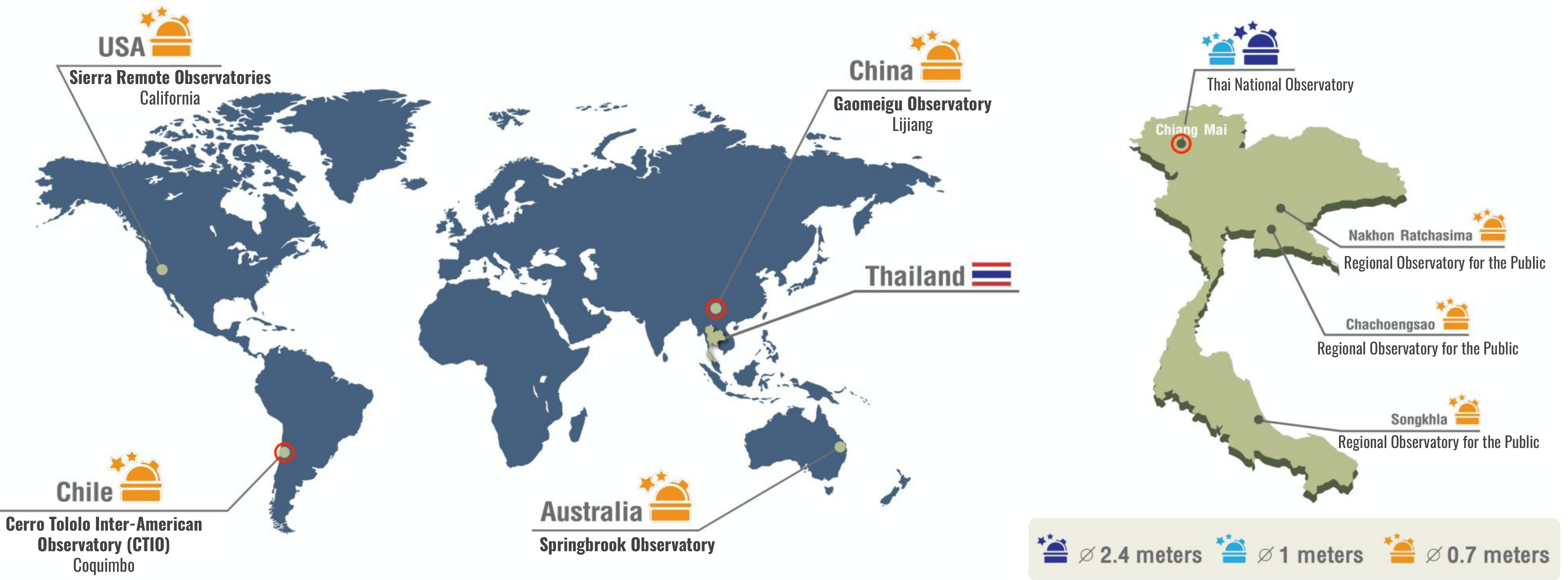}
    \caption{The location and size of the telescopes currently within the SPEARNET network. Telescopes with a red circle are those used to take the observations of WASP-127~b discussed in Section \ref{sec:WASP-127b}. (Image credit: NARIT)}
    \label{fig:SPEARNET_telescopes}
\end{figure*}

SPEARNET is a prototype transmission spectroscopy survey which is utilising a globally-distributed network of heterogeneous optical telescopes, the locations of which are shown in Figure \ref{fig:SPEARNET_telescopes}. It was conceived to anticipate and address the challenges that the transition into the so-called ``asset-starved'' era poses, primarily by designing tools which allow for increased utility of resources, both before \citep{morgan2019metric} and after \citep{Hayes2020} transit observations. \tranf{} was conceived as part of the SPEARNET suite of tools to handle  transit data from non-homogeneous observations, to facilitate transmission spectroscopy studies in the asset-staved era, as time on larger telescopes is becoming ever-more competitive and studies will have to frequently rely on data taken from a combination of telescopes. 

As part of the network operation, \citet{morgan2019metric} developed a metric for ranking candidates for observation, effectively pairing targets with telescopes in a way which maximises the signal-to-noise of the observations. The motivation behind this metric is to remove the multiple unquantifiable biases in manual transmission spectroscopy target selection. Since the selection function is known, it is possible to make population-corrected statistical statements based on observations that are guided by the metric. In Table 3 of \citet{morgan2019metric}, we show that WASP-127~b is consistently ranked in the top three targets for a variety of telescopes when known planet masses are included in the metric calculations, and as such it has become a target of interest for SPEARNET.

With a density of $0.07\pm 0.01~\rho_\text{Jup}$ \citep{WASP-127b_discovery}, WASP-127~b is one of the lowest-density planets so far discovered, and occupies the `short-period Neptune desert' \citep{neptune_desert}, which is notable since most planets with its characteristics are not expected to survive due to photo-evaporation \citep{photoevaporation}. Its low density also makes WASP-127~b an idea target for transmission spectroscopy due to its large atmospheric scale height, and several studies have been completed, with potential detections of water \citep{WASP-127b_NaK&Li, WASP-127b_2020b}, and statistically significant detections of sodium, potassium, and lithium \citep[$5\sigma$, $3\sigma$, and $4\sigma$ respectively,][]{WASP-127b_NaK&Li}. No significant evidence for helium in the upper atmosphere has been found \citep{WASP-127b_2020a} and it has been proposed that this is due to unfavourable photo-ionisation conditions. 

The approaches to LDC fitting in these previous studies differ. \citet{WASP-127b_2020b} fix the LDCs for all spectral channels at the white-light values predicted for WASP-127 using the quadratic law of \citet{Claret2000}. \citet{WASP-127b_NaK&Li} also use a quadratic limb-darkening law, but instead find the highest likelihood values for each channel and fit using a Gaussian prior of width $0.1$, sourcing the initial predictions from the Kurucz ATLAS9 stellar atmosphere models \citep{ATLAS9}.

Using the SPEARNET telescope network, we have obtained six photometric light curves in four different wavebands from five transits of WASP-127~b, including the first published transits observed in the $u'$-band. The three telescopes used in these observations were:
\begin{description}
    \item \emph{The 2.4m Thai National Telescope (TNT)}:
    \newline\indent Located at the Thai National Observatory (TNO), the TNT observations of WASP-127b were conducted using ULTRASPEC \citep{Dhillon2014}, which uses a $1024\times1024$~pixel high-speed frame-transfer EMCCD camera with a field-of-view of $7.68 \times 7.68$~arcmin$^{2}$. The dead time between exposures on this setup is $14$~ms.
    
    \item \emph{A 0.7~m telescope at Gao Mei Gu observatory (TRT-GAO)}
    \newline \indent The TRT-GAO is also part of the Thai Robotic Telescope Network, and is located at Gao Mei Gu observatory in Lijiang, China. Observations were taken using an Andor iKon-L 936 $2048\times2048$ CCD with a field of view of $20.9\times20.9$~arcmin$^2$.
    
    \item \emph{The 0.6~m PROMPT-8 telescope}
    \newline \indent Located at the Cerro Tololo Inter-American Observatory (CTIO) in Chile, PROMPT-8 is a $0.6$~m telescope operated by the Skynet Robotic Telescope Network. Imaging is conducted on this telescope using a $2048\times2048$~pixel CCD camera with a scale of $0.624$~arcseconds/pixel.
\end{description}

Figure \ref{fig:SPEARNET_telescopes} shows the location of the primary telescopes in the SPEARNET network, with the telescopes used in this study of WASP-127~b highlighted. The precise details of the observations taken are given in Table \ref{tab:WASP-127b_observations}.

\begin{table*}
    \centering
    \caption{Details of the SPEARNET observations of WASP-127~b, obtained using the Thai National Telescope (TNT), telescope from the Thai Robotic Telescope network at Gao Mei Gu (TRT-GAO), and the PROMPT-8 telescope at the Cerro Tololo Inter-American Observatory. Since the observations were taken, TRT-TNO has been upgraded to a $1$~m aperture telescope.}
    \label{tab:WASP-127b_observations}
    \begin{tabular}{c c c c c c c}
        \hline \hline
        Date & Telescope & Filter & Aperture (m) & Number of photometric points & Exposure time (s) & Transit coverage \\
        \hline
        15-02-2017 & TNT & $i'$ & $2.4$ & 24 720 & 0.78 & Full \\
        15-02-2017 & TRT-GAO & $R$ & $0.7$ & 1 223 & 15, 20 & Full \\
        21-03-2017 & PROMPT-8 & $R$ & $0.6$ & 413 & 5 & Egress only\\
        26-02-2018 & TNT & $r'$ & $2.4$ & 68 480 & 0.38 & Full\\
        09-03-2019 & TNT & $u'$ & $2.4$ & 741 & 12.8 & Ingress only\\
        30-03-2019 & TNT & $u'$ & $2.4$ & 1 531 & 12.8 & Full\\
        \hline
    \end{tabular}
\end{table*}

The images obtained by SPEARNET were calibrated using \texttt{IRAF} routines along with astrometric calibration using Astrometry.net \citep{Lang2010}. In order to obtain the light curve, aperture photometry was carried out using \texttt{sextractor} \citep{Bertin1996} using an adaptive scaled aperture based on the seeing in an individual image. Reference stars were chosen to have magnitudes similar to WASP-127 ($\left | \Delta M \right | < 3$) and no intrinsic variation. The observed time in BJD$_{\textrm{TDB}}$ with flux ratio between WASP-127 b and reference stars with their error are shown in Table \ref{tab:WASP-127b_LC}.

\begin{table}
	\centering
 	\caption{The photometric data of WASP-127 b.}
	\label{tab:WASP-127b_LC}
	\begin{tabular}{ccccc}
		\hline \hline
		BJD$_{\textrm{TDB}}$ & Flux & Flux error & Filters & Telescopes \\
		\hline
        2 457 800.1182452 & 2.387 & 0.007 & $i'$ & TNT \\
        2 457 800.1182542 & 2.424 & 0.008 & $i'$ & TNT \\
        2 457 800.1182633 & 2.367 & 0.007 & $i'$ & TNT \\
        2 457 800.1182724 & 2.394 & 0.008 & $i'$ & TNT \\
        2 457 800.1182815 & 2.379 & 0.007 & $i'$ & TNT \\
        2 457 800.1182905 & 2.358 & 0.007 & $i'$ & TNT \\
        2 457 800.1182996 & 2.414 & 0.008 & $i'$ & TNT \\
        2 457 800.1183087 & 2.430 & 0.008 & $i'$ & TNT \\
        2 457 800.1183178 & 2.408 & 0.007 & $i'$ & TNT \\
        \vdots & \vdots & \vdots & \vdots & \vdots \\
        \\
		\hline
	\end{tabular}
\\ Note. The raw lightcurves are available at the CDS.\protect\footnotemark%\footnote{\url{https://cdsarc.u-strasbg.fr/ftp/vizier.submit/transitfit/}}}.

\end{table}
\footnotetext{\url{https://cdsarc.u-strasbg.fr/ftp/vizier.submit/transitfit_data/}}%

\begin{figure*}
    \centering
    \includegraphics[width=\textwidth]{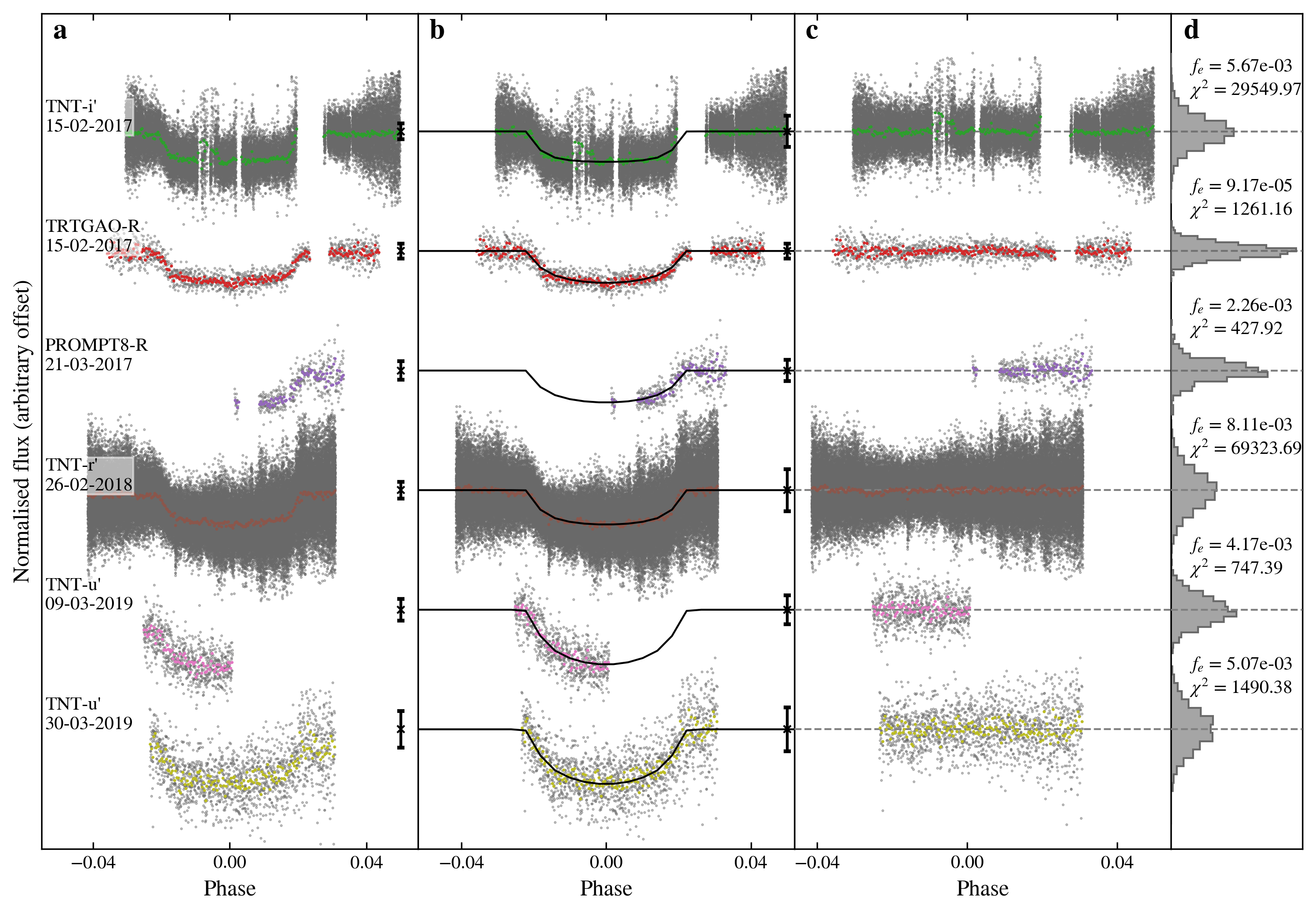}
    \caption{(\textbf{a)}: The raw data from the SPEARNET observations of WASP-127~b outlined in Table \ref{tab:WASP-127b_observations}, normalised using the median values for each observation and phase folded using the model obtained in (\textbf{b}). We have binned the observations to a 2-minute cadence, with the raw observations shown in grey. For clarity, error bars on data points have been excluded and an average error bar of the raw data points has been provided as a black coloured bar to the right of each curve. (\textbf{b}): The detrended light curves and associated best-fit models obtained using \tranf{} in ``batched'' mode with a coupled quadratic LDC fitting and simultaneous normalisation and 2nd-order detrending. Each curve has an arbitrary offset from a normalised baseline of $1$ and has been phase-folded to centre $t_0$ at a phase of 0.0. The best fit transit model from the retrieval is over-plotted. We show the average of the errors scaled using retrieved value of $f_e$, as a black coloured point on the right of each curve. (\textbf{c}): The residuals after model is subtracted from the fitted lightcurves. (\textbf{d}): The histogram of the residuals along with the respective $\chi^2$ values.}
    \label{fig:WASP-127b_fitted}
\end{figure*}

The obtained light curves, shown in \ref{fig:WASP-127b_fitted}(a), were then run through \tranf{}, using the `batched' mode, a 2nd-order detrending function, using quadratic limb-darkening model, and both `coupled' and `independent' LDC fitting approaches so as to able to identify the improvement from using filter and host parameters to inform the likelihood of LDC values. For the coupled LDC approach, we adopted the stellar parameters of \citet{WASP-127b_discovery}, namely $T=5620\pm85$~K, $R_\star=1.39\pm0.03~R_\odot$, $M_\star=1.08\pm0.03~M_\odot$, $\log g=4.18\pm0.01$~cgs, and [Fe/H]~$=-0.18\pm0.06$~dex. For the priors, we take the values from \citet{WASP-127b_NaK&Li}. The value of $t_0$ was scaled using $P$ to match the time span of the raw lightcurves, to reduce extrapolation during fitting.

\begin{figure}
    \centering
    \includegraphics[width=\columnwidth]{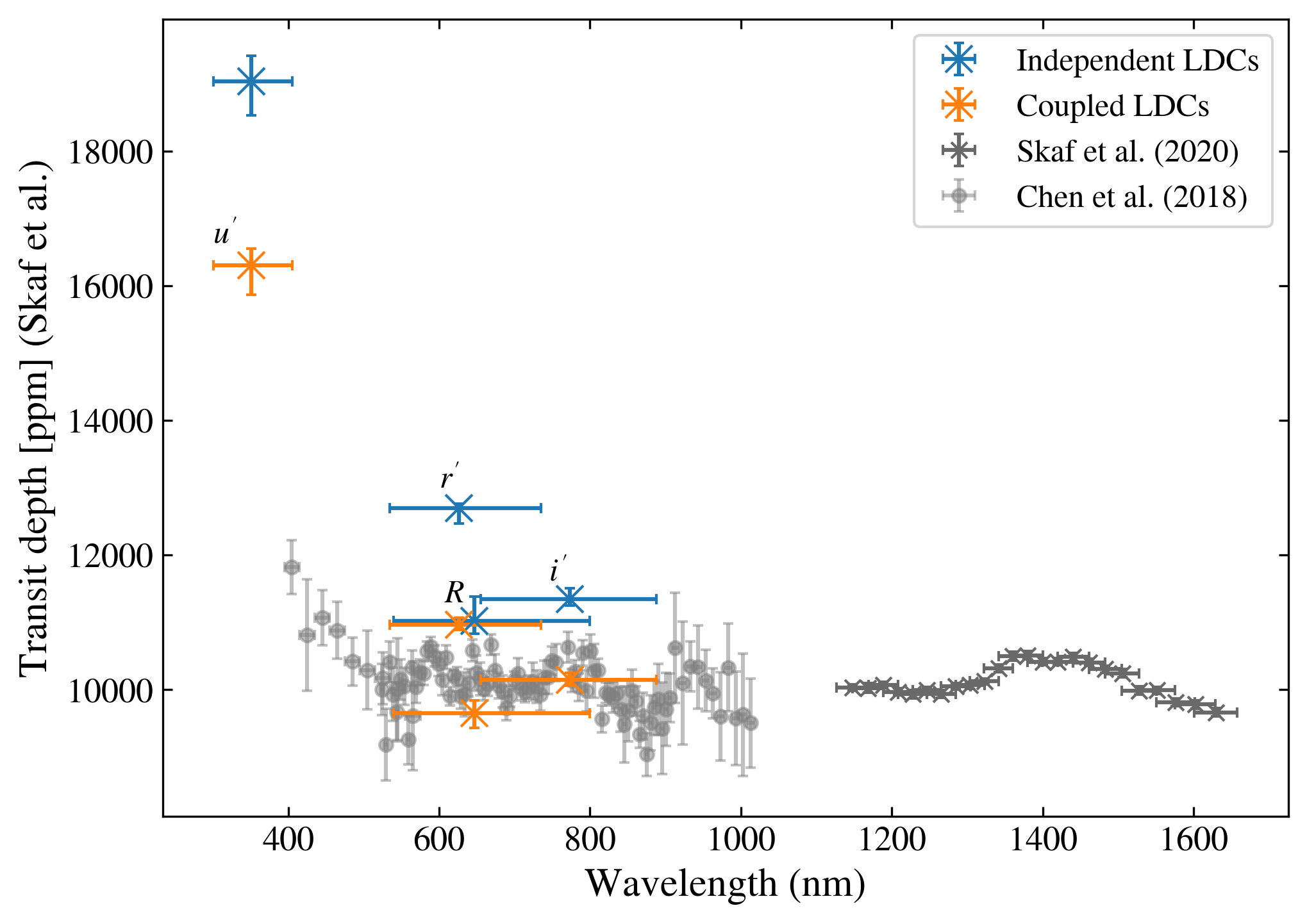}
    \caption{The transmission spectrum of WASP-127~b, obtained by fitting ground-based observations from the SPEARNET telescope network using \tranf{} in coupled (orange) and independent (blue) LDC fitting modes with a second-order detrending function. From left to right the filters are $u'$,  $r'$, $R$, and $i'$, and we have labelled the independent LDC depths for quick reference.  Also shown are the GTC/NOT observations (after applying corrective offset) reported in \citet[][light grey]{WASP-127b_NaK&Li} and the \emph{HST} reported in \citet[][dark grey]{WASP-127b_2020b}. For the SPEARNET data points, the horizontal positions and error bars are derived from the SED-weighted filter profiles, with the data point positioned at the weighted mean of each profile and the error bars reflecting the standard deviations. For the other data sources, horizontal error bars represent the complete bandwidth of each spectroscopic filter.}
    \label{fig:WASP-127b_spectrum}
\end{figure}

The resulting detrended light curves from the coupled LDC fitting are shown with the best-fit model in Figure \ref{fig:WASP-127b_fitted}(b). The corresponding corner plot showing posteriors are shown in Figure \ref{fig:posterior-plot} in Appendix \ref{sec:posterior_plot}. We also show the resulting transit depths from both the coupled and independent LDC mode retrievals alongside the \textit{Hubble Space Telescope} (\textit{HST}) spectrum from \cite{WASP-127b_2020b}, and the Gran Teliscopio Canarais (GTC) and Nordic Optical Telescope (NOT) spectral data and best fit model from \cite{WASP-127b_NaK&Li} in Figure \ref{fig:WASP-127b_spectrum}. 

Since our data, and those of \citet{WASP-127b_NaK&Li} and \cite{WASP-127b_2020b} have all been analysed separately, there is an intrinsic offset between all the data. 
We have compared the data with our retrieved spectrum in Figure \ref{fig:WASP-127b_spectrum}.
The wavelength positions of the SPEARNET observations in Figure \ref{fig:WASP-127b_spectrum} are derived by weighting the relevant filter profile by the spectral energy distribution (SED) of WASP-127, predicted using the PHOENIX models.

Looking at Figure \ref{fig:WASP-127b_spectrum}, from coupled LDC fitting approach, the value of $R_p/R_\star$ is somewhat different than the independent LDC fitting method, with a discrepancy of around $8$~per~cent in the $u'$ and $r$ bands. This difference is larger than the $3$~per~cent bias that \cite{LDC_fitting_in_exoplanets} find can be introduced from not fitting LDCs at all, and clearly illustrates the impact that using host characteristics and filter profiles can have when fitting spectroscopic and photometric measurements.

This could also be due to the incompleteness of the observation, as in the case of $R$-band and $u'$-band, which do not have a complete transit observed. When using the coupled LDC approach, the fact that this observation does not have a complete transit may have an effect on the parameters retrieved for all the filters. Table \ref{tab:WASP-127b_results} shows the physical results obtained by each of the two retrievals described above.

Looking at Figure \ref{fig:WASP-127b_spectrum}, it may be noticed that the transit depth precision obtained from the TNT observations are roughly comparable to those obtained by \citet{WASP-127b_2020b} from their \emph{HST} observations. At first glance, this may appear surprising, as one would instinctively expect that space-based observations would produce higher-quality results than those made from ground-based observatories. However, this fails to account for the impact of the high on-sky efficiency of observations made using ULTRASPEC as well as the difference between broadband imaging and spectroscopy. Assuming that calculations are photon-noise limited, we expect the 
signal-to-noise ratio (S/N) on the retrieved flux value to go as $\sqrt{N_\gamma}$, where $N_\gamma$ is the number of photons collected. Using this, S/N for the transit depth can be estimated by $\sqrt{N_{\gamma,\textrm{t}_{14}}+N_{\gamma,\textrm{base}}}$ where $N_{\gamma,\textrm{t}_{14}}$ and $N_{\gamma,\textrm{base}}$ are the number of photons in-transit and in baseline respectively. In the source-dominated limit $N_\gamma$ can be estimated by
\begin{equation}
    N_\gamma = t_\textrm{int} 10^{-0.4\left(m - m_0\right)},
    \label{eq:N_photons}
\end{equation} 
where $t_\textrm{int}$ is the on-sky integration time, $m$ is the host apparent magnitude and $m_0$ the instrument zero-point magnitude.

For TNT in $r'$ the zero-point magnitude is $m_{0,{\rm TNT}} = 25.25$\footnote{\url{http://deneb.astro.warwick.ac.uk/phsaap/ULTRASPEC/calibration.html}} and with 42,680 in-transit observations with a mean exposure of 0.38~s this gives $N_{\gamma,\text{t}_{14},\text{TNT}} \simeq 1.7 \times 10^{10}$ integrated over all in-transit exposures of WASP-127 ($r' = 10.0$). With 26,088 baseline observations, we also get $N_{\gamma,\text{base},\text{TNT}} \simeq 1.0 \times 10^{10}$.

\cite{WASP-127b_2020b} observed 38 in-transit observations using the \emph{HST} Wide Field Camera 3 (WFC3) G141 grism with a mean exposure of 95.782~s. Using the WFC3IR spectroscopic exposure time calculator\footnote{\url{https://etc.stsci.edu/etc/input/wfc3ir/spectroscopic/}}, and adopting a G5V host spectrum for WASP-127 normalised to $V = 10.15$~mag, we find a photon count integrated over all in-transit exposures of $N_{\gamma,\text{t}_{14},\text{HST}} \simeq 3.4 \times 10^{9}$ at 1.4~$\mu$m for a spectral resolution of 70. The corresponding baseline count gives us $N_{\gamma,\text{base},\text{HST}} \simeq 3.2 \times 10^{9}$. This gives a rough ratio of S/N for \emph{TNT} to S/N for \emph{HST}, of
$$\sqrt{\frac{N_{\gamma,\text{t}_{14},\text{TNT}}+N_{\gamma,\text{base},\text{TNT}}}{N_{\gamma,\text{t}_{14},\text{HST}}+N_{\gamma,\text{base},\text{HST}}}} \simeq 4.2.$$ 
Clearly, the approximate equality here masks the fact that we are comparing the throughput of a broadband filter on TNT to the sensitivity of a single spectral bin from the G141 grism.

\begin{table*}
    \centering
    \caption{Output parameters from using \tranf{} to run retrieval on photometric observations of WASP-127b, in both coupled and independent LDC fitting mode. For comparison, we provide the orbital parameters obtained by \protect{\citet{WASP-127b_NaK&Li}} and \protect{\citet{WASP-127b_2020b}}. We have converted the \citet{WASP-127b_NaK&Li} value of $a/R_\star$ into AU using the \citet{WASP-127b_discovery} value of $R_\star=1.39\pm0.03~R_\odot$. }
    \label{tab:WASP-127b_results}
    \begin{tabular}{c c c}
        \hline\hline
        & \multicolumn{2}{c}{SPEARNET observations}\\
         & Coupled LDC fitting & Independent LDC fitting \\
         \hline\\[-1em]
        $P$~[days] & $4.1780626^{+ 0.0000019}_{-0.0000026}$ & $4.1780667^{+ 0.0000015}_{-0.0000019}$ \\[.35em]
[$\text{BJD}$$_{\textrm{TDB}}$]  & $2457800.24581^{+ 0.00016}_{-0.00018}$ & $2457800.24543^{+ 0.00021}_{-0.00009}$ \\[.35em]
$i$~[deg]  & $89.64^{+ 0.24}_{-0.20}$ & $86.36^{+ 0.28}_{-0.12}$ \\[.35em]
$a/R_\star$  & $7.837^{+ 0.026}_{-0.016}$ & $7.333^{+ 0.117}_{-0.051}$ \\[.35em]
$a$~[AU]  & $0.0507^{+0.0011}_{-0.0011}$ & $0.0474^{+0.0013}_{-0.0011}$ \\[.35em]
$R_p/R_\star$~[$u'$- band]  & $0.1277^{+ 0.0010}_{-0.0017}$ & $0.1380^{+ 0.0014}_{-0.0018}$ \\[.35em]
$R_p/R_\star$~[$r'$- band]  & $0.1047^{+ 0.0005}_{-0.0003}$ & $0.1127^{+ 0.0003}_{-0.0010}$ \\[.35em]
$R_p/R_\star$~[$R$- band]  & $0.0983^{+ 0.0010}_{-0.0011}$ & $0.1050^{+ 0.0017}_{-0.0009}$ \\[.35em]
$R_p/R_\star$~[$i'$- band]  & $0.1007^{+ 0.0005}_{-0.0004}$ & $0.1065^{+ 0.0008}_{-0.0004}$ \\[.35em]
$u_0$~[$u'$-band]  & $0.868^{+0.078}_{-0.036}$ & $0.934^{+0.103}_{-0.147}$ \\[.35em]
$u_1$~[$u'$-band]  & $0.010^{+0.077}_{-0.029}$ & $-0.446^{+0.191}_{-0.165}$ \\[.35em]
$u_0$~[$r'$-band]  & $0.534^{+0.023}_{-0.036}$ & $0.709^{+0.034}_{-0.055}$ \\[.35em]
$u_1$~[$r'$-band]  & $0.363^{+0.033}_{-0.041}$ & $-0.350^{+0.066}_{-0.063}$ \\[.35em]
$u_0$~[$R$-band]  & $0.763^{+0.061}_{-0.042}$ & $0.514^{+0.275}_{-0.154}$ \\[.35em]
$u_1$~[$R$-band]  & $0.143^{+0.060}_{-0.039}$ & $0.001^{+0.272}_{-0.131}$ \\[.35em]
$u_0$~[$i'$-band]  & $0.287^{+0.034}_{-0.026}$ & $0.102^{+0.124}_{-0.053}$ \\[.35em]
$u_1$~[$i'$-band]  & $0.646^{+0.057}_{-0.041}$ & $0.271^{+0.157}_{-0.139}$ \\[.35em]
        \hline\hline
        & \multicolumn{1}{c}{\protect{\citet{WASP-127b_NaK&Li}}} & \multicolumn{1}{c}{\protect{\citet{WASP-127b_2020b}}}\\
        \hline
        $P$~[days]& \multicolumn{1}{c}{$4.17807015\pm5.7\times10^{-7}$} & \multicolumn{1}{c}{$4.1780619\pm1.3\times10^{−6}$}\\
        $t_0$~[$\text{BJD}$$_{\textrm{TDB}}$]& \multicolumn{1}{c}{$2457248.741276\pm0.000068$} & \multicolumn{1}{c}{$2458238.943367\pm0.000055$}\\
        $i$~[deg]& \multicolumn{1}{c}{$87.88\pm0.32$} & \multicolumn{1}{c}{$88.2\pm1.1$}\\
        $a/R_\star$& \multicolumn{1}{c}{$7.846\pm0.089$} & \multicolumn{1}{c}{$7.846~^{\textrm{a}}$}\\
        $a$~[AU]& \multicolumn{1}{c}{$0.0507\pm0.0012$} & \multicolumn{1}{c}{$0.0507~^{\textrm{a}}$}\\
        \hline
        \multicolumn{3}{l}{$^\textrm{a}$ \protect{\citet{WASP-127b_2020b}} adopt the value of $a/R_\star$ directly from \protect{\citet{WASP-127b_NaK&Li}}}
    \end{tabular}
\end{table*}

\subsection{Multi-epoch study of WASP-91~b with \textit{TESS}}
\label{sec:WASP-91b}
WASP-91~b is a $1.34~\mjup$ warm Jupiter with a 2.8~day orbit around a K3 host star \citep{WASP-91b_discovery}. With a radius of $1.03~\rjup$, WASP-91~b is a smaller example of a hot Jupiter, and has not been the subject of further studies since its discovery. 

\textit{TESS} observed WASP-91~b in Sectors 1, 27, and 28, capturing a total of 50\,158 photometric data points covering 26 transits, the data which we acquired  from the Barbara A. Mikulski Archive for Space Telescopes (MAST) portal\footnote{\url{https://mast.stsci.edu}}. We use 120~s cadence data, and PDC$\_$SAPFLUX values are taken as flux. Since these data contain vast numbers of observations outside of transit, and there are multiple offsets and other trends within the data which cannot be modelled with a simple polynomial, we estimate the transit duration $t_{14}$, assuming a circular orbit with a $90$~degree inclination, using 
\begin{equation}
    t_{14} = \frac{R_p + R_\star}{a\pi}P
    \label{eq:t14}
\end{equation}
and discard any data which are more than $2.5t_{14}$ away from the mid-transit times predicted by the ephemeris given in \citet{WASP-91b_discovery}. This results in 26 transits, capturing a combined total of 9\,789 photometric points, which can then be individually normalised and detrended with low-order polynomials. This splitting of data is provided in \tranf{} through the \texttt{split\_lightcurve\_file} function.

After splitting the full data into individual transit observations, we use the ``folded'' mode of \tranf{} to fit these data, using both the ``coupled'' and ``independent'' LDC fitting modes, assuming a circular orbit and using the quadratic limb-darkening model from Equation \ref{eq:quadraticLD} and a second-order detrending polynomial. The values from \citet{WASP-91b_discovery} were used as priors for our fitting. For the ``coupled'' mode, we inform LDC fitting using the stellar parameters found by \citet{WASP-91b_discovery}, given in Table \ref{tab:WASP_91b_results}, and the \textit{TESS} filter profile given on by the SVO Filter Profile Service\footnote{\url{http://svo2.cab.inta-csic.es/theory/fps/}} \citep{filter_profile_service1, filter_profile_service2}. 

\begin{figure}
    \centering
    \includegraphics[width=\columnwidth]{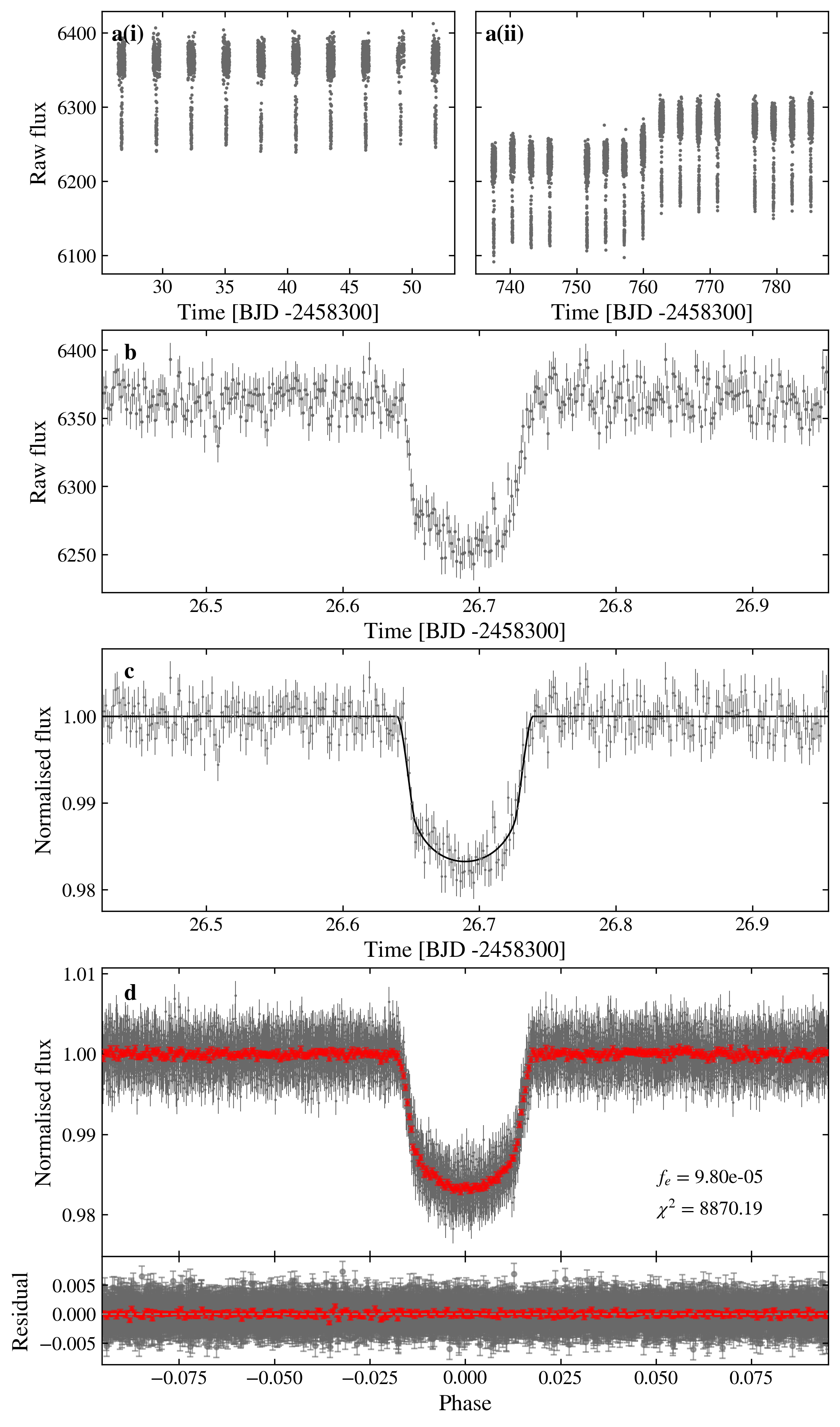}
     \caption{The full data processing of the \textit{TESS} observations of WASP-91~b using the ``folded'' retrieval mode of \tranf{}, with LDCs informed by LDTk. \textbf{a(i)}: the raw observations from \textit{TESS} Sector 1; \textbf{a(ii)}: the raw observations from Sectors 27 and 28. We have excluded the error bars from these two plots for clarity; \textbf{b}: The raw data from the first observed transit, chosen purely as an illustration; \textbf{c}: The first observed transit after the first stage of ``folded'' mode retrieval. These data have been detrended and normalised, the error bars have been scaled using the retrieved $f_e$, and have the final best-fit model overplotted; \textbf{d}: the final phase-folded light curve, best-fit transit model and residuals, containing 9\,789 photometric data points, and, in red, the same data binned to a 2 minute observation cadence. The error bars have been scaled using $f_e$, and the corresponding $\chi^2$ value has also been displayed.}
    \label{fig:WASP-91b_results}
\end{figure}

Figure \ref{fig:WASP-91b_results} shows the fitting process for the ``coupled'' run of \tranf{} at various stages. Figure \ref{fig:WASP-91b_results} a(i) and a(ii) show the raw \textit{TESS} observations which clearly exhibit various offsets and long-term trends. In order to reduce the impact of these, we split the data into individual transits using the approach described above, and Figure \ref{fig:WASP-91b_results}(b) shows the resulting raw data for the first transit. In the ``folded'' fitting mode, each transit is normalised and detrended, and Figure \ref{fig:WASP-91b_results}(c) shows the first transit after this first stage of processing. We have overplotted the final best fit model to help illustrate this step, but it should be noted that this model is calculated from all the transits, not just this single epoch. Once normalisation and detrending has been fitted for each transit, all the light curves are folded together and this curves is analysed to retrieve the final best-fit model. Figure \ref{fig:WASP-91b_results}(d) shows this folded light curve, along with the final best-fit transit model and residuals, which have an rms of 0.0017. We also show the same folded data binned to a cadence of two minutes, to demonstrate the improvement in observation precision when compared to the single-transit \textit{TESS} observations like the one shown in Figure \ref{fig:WASP-91b_results}(c). An average of nearly $25$ photometric points are contained within each of the bins, which through na\"ive root-$N$ statistics suggests a maximum improvement in precision of factor $5$. The rms of the residuals of this binned light curve is 0.0004, which is an improvement of factor 4.8. The similarity of these two factors suggest that the improvement due to folding is near maximal, and thus the binned data are not systematics-limited.

We present the results from both runs of \tranf{} alongside the results from \citet{WASP-91b_discovery} in Table \ref{tab:WASP_91b_results}, which are all consistent with each other. 
Precise ephemerides are required for accurate study of TTVs, and updating ephemeris by applying \tranf{} to planets within \textit{TESS} data will prove invaluable to future surveys.

The uncertainties on the two \tranf{} retrievals are generally comparable, with the notable exception of the LDCs. Through using LDTk to calculate LDC likelihoods, we see upto an order-of-magnitude increase in the precision of LDC values, which demonstrates the impact of introducing host parameters and filter information into transit-fitting routines. 

\begin{table*}
    \centering
    \caption{Output parameters for retrieval using \tranf{} in ``folded'' retrieval mode on \textit{TESS} observations of 26 transits of WASP-91~b, using both ``coupled'' and ``uncoupled'' LDC fitting approaches, alongside those obtained by \protect\citet{WASP-91b_discovery}. The data were fitted using a 2nd-order detrending model and the quadratic limb-darkening law given in Equation \ref{eq:quadraticLD}, and assuming zero orbital ellipticity, using the profile of the \emph{TESS} filter available on the SVO Filter Profile Service to calculate LDC likelihoods when using LDTk. The \tranf{} retrieval uses host parameters fixed at those obtained by \protect\cite{WASP-91b_discovery}, which are used in tandem with the filter profile to inform the LDC likelihood fitting.}
    \label{tab:WASP_91b_results}
    \begin{tabular}{c c c c}
        \hline \hline
        & \tranf{}: coupled LDCs & \tranf{}: independent LDCs & \citet{WASP-91b_discovery}\\
        \hline\\[-1em]
        $P$~[days] & $2.7985783^{+ 0.0000024}_{-0.0000027}$ & $2.7985794^{+ 0.0000021}_{-0.0000020}$  & $2.798581 \pm 3\times10^{-6}$ \\[.3em]
        $t_0$~[$\text{BJD}$$_{\textrm{TDB}}$]  & $2458326.6895^{+ 0.0006}_{-0.0004}$ & $2458326.6894^{+ 0.0005}_{-0.0005}$  & $2456297.7190 \pm 0.0002$~$^\textrm{a}$\\[.3em]
        $i$~[deg]  & $87.36^{+ 0.20}_{-0.17}$ & $87.21^{+ 0.31}_{-0.17}$  &  $86.8\pm 0.4$\\[.3em]
        $a/R_\star$  & $9.34^{+ 0.12}_{-0.11}$ & $9.28^{+ 0.16}_{-0.10}$  &  $9.251\pm 0.408$\\[.3em]
        $a$~[AU]  & $0.0374^{+0.0014}_{-0.0014}$ & $0.0371^{+0.0014}_{-0.0014}$  &  $0.037\pm0.001$\\[.3em]
        $R_p/R_\star$ & $0.1198^{+ 0.0005}_{-0.0005}$ & $0.1206^{+ 0.0008}_{-0.0013}$  &  $0.1225 \pm 0.0012$\\[.3em]
        $R_p$~[$\rjup$] & $1.002^{+ 0.039}_{-0.031}$ & $1.009^{+ 0.042}_{ -0.024}$  &  $1.03 \pm 0.04 $\\[.3em]
        $u_0$ &  $0.505^{+ 0.021}_{-0.031}$ & $0.551^{+ 0.137}_{-0.210}$ & -~$^\textrm{b}$\\[.3em]
        $u_1$ &  $0.121^{+ 0.021}_{-0.031}$ & $0.016^{+ 0.090}_{-0.203}$ & -~$^\textrm{b}$\\[.3em]
        $T_{\text{eff},\star}$~[K] &- & -& $4920\pm80$\\
        $M_\star$~[$M_\odot$] & -&- &$0.84 \pm 0.07$ \\
        $R_\star$~[$R_\odot$] &-&- &$0.86 \pm 0.03$\\
        ~[Fe/H] &-&- &$0.19 \pm 0.13$ \\
        \hline
        \multicolumn{4}{l}{$^\textrm{a}$ The time standard is not specified}.\\
        \multicolumn{4}{l}{$^\textrm{b}$ \citet{WASP-91b_discovery} use the non-linear limb-darkening law but do not provide coefficients to compare with.}
    \end{tabular}
\end{table*}

\subsection{TTV analysis of WASP-126~b}
\label{sec:WASP-126b}
Orbiting a type G2 star with a period of $3.2888\pm0.0008$~days, WASP-126~b \citep{WASP-126b_discovery} is a $0.28\pm0.04~\mjup$ hot Jupiter which has been identified as potentially exhibiting TTVs. Through Bayesian $N$-body simulation coupled with machine learning analysis of Sectors 1--3 of the \textit{TESS} observations of WASP-126~b, \citet{WASP-126b_TTVs} showed that there was evidence of a TTV signal with amplitude of $\sim1$~minute and a period of $\sim25$~days, which could be attributed to a non-transiting planet with $M_p = 0.202\pm0.077~\mjup$ on a $7.63\pm0.17$~day orbit, dubbed WASP-126~c. \citet{WASP-126c_no-evidence} studied the \textit{TESS} observations from sectors 1--13 and found that when the extra sectors were included, the TTV signal was not present at a statistically significant level. 

Here we use the ability of \tranf{} to account for TTVs to analyse the \textit{TESS} observations of WASP-126~b from sectors 1--13, 27--34, 36--39, 61, and 63 to further investigate the presence of TTVs within these data, and to produce the most up-to-date values for the planetary and orbital parameters of WASP-126~b. As with the analysis of WASP-91~b, we use PDC$\_$SAPFLUX values for flux, and we discard all data that is more than $2.5t_{14}$ from the mid-transit times predicted by \citet{WASP-126b_discovery}, which gives 180 individual transits. We then run analysis using \tranf{} in both ``coupled'' and ``independent'' LDC fitting modes, using the quadratic limb-darkening law, a 2nd order detrending polynomial, assuming a circular orbit, and using results from \citet{WASP-126b_discovery} as priors. When using LDTk to calculate LDC likelihoods, we use the host parameters given in \citet{WASP-126b_discovery}, which are presented in Table \ref{tab:WASP-126b_results}, and the \textit{TESS} filter profile from the SVO Filter Profile Service. As discussed in Section \ref{sec:ttvs}, we first run analysis of the data assuming that there are no TTVs, and we use these results to provide priors to parameterise the ephemeris for the analysis where we allow for the presence of TTVs. Since \tranf{} requires a fixed period to be provided when considering TTVs, this step should always be used to ensure complete consistency of results. The priors used in this initial step are based on the ephemeris of \citet{WASP-126b_discovery}. The resulting ephemerides from this initial analysis are given in Table \ref{tab:WASP-126b_results}, and we use these results when allowing for the presence of TTVs, fixing the period at the values given and using the retrieved $t_0$ values as the mean of a Gaussian prior with a width of $0.007$~days.

The orbital parameters of WASP-126~b retrieved by \tranf{} when allowing for TTVs are given in Table \ref{tab:WASP-126b_results}, for both ``coupled'' and ``independent'' LDC fitting modes. We present these alongside the results of \citet{WASP-126b_TTVs}, \citet{WASP-126c_no-evidence}, and \citet{WASP-126b_discovery} for comparison. We find that the results from both runs are generally consistent but, as in the WASP-91~b analysis, the uncertainties on the LDCs for the ``coupled'' run are smaller. 

\begin{figure*}
    \centering
    \includegraphics[width=\textwidth]{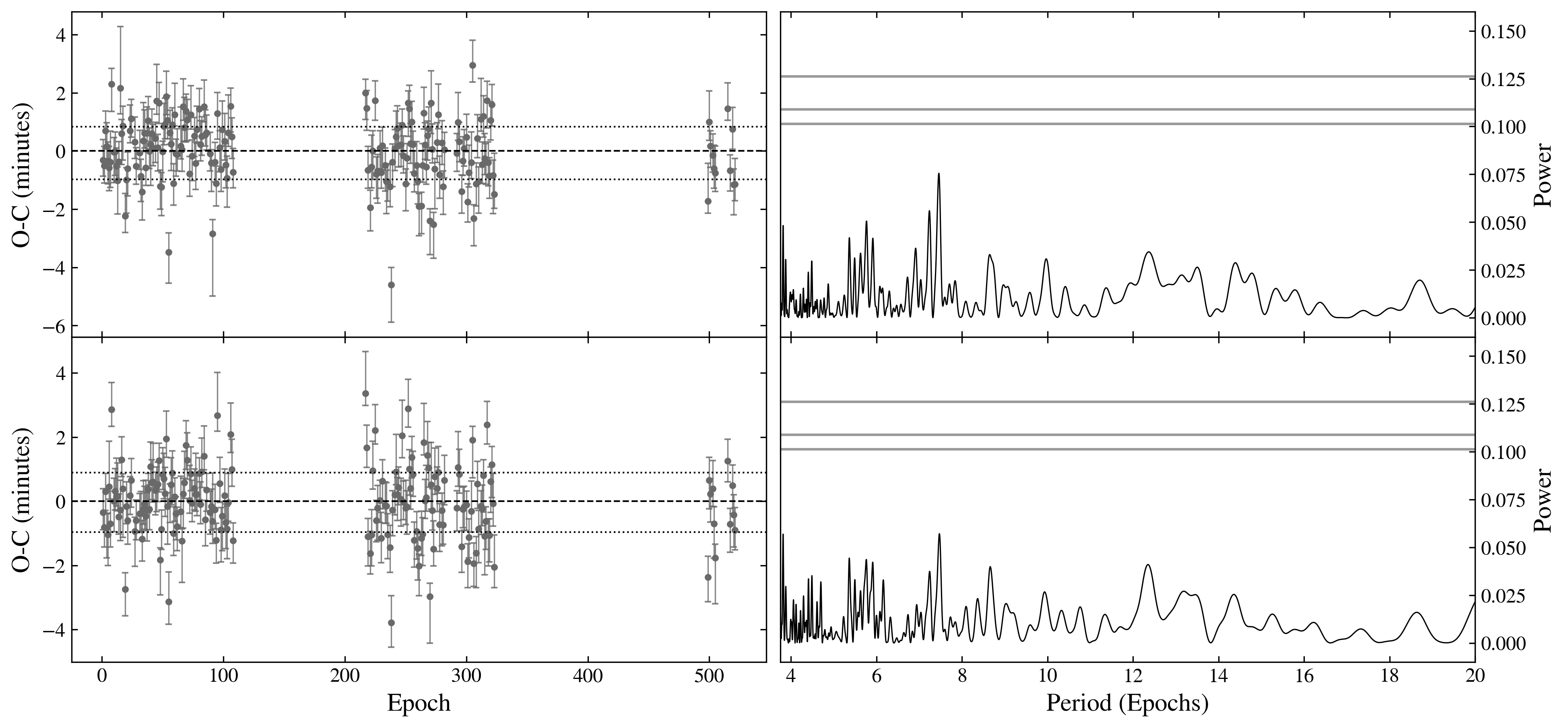}
    \caption{\textit{Left}: The O-C plots of 126 transits of WASP-126~b observed by \textit{TESS}, obtained from results of using \tranf{} in ``coupled'' (top) and ``independent'' (bottom) LDC modes whist allowing for the presence of TTVs. The values of $t_0$ in this case had a median error of 4.97$\times10^{-4}$ in the coupled mode and 4.90$\times10^{-4}$ in the independent mode. The dotted lines represent the uncertainty on the predicted ephemerides derived from using \tranf{} without allowing for TTVs. \textit{Right}: The associated weighted Lomb-Scargle periodograms for each of the O-C plots. The horizontal grey lines represent the false alarm probabilities of $10$, $5$, and $1$~per~cent from bottom to top.}
    \label{fig:WASP-126b_O-C}
\end{figure*}

We present the O-C plots for both modes of the \tranf{} analysis in the left plots of Figure \ref{fig:WASP-126b_O-C}, with the associated Lomb-Scargle periodograms on the right. The top row are the results for the ``coupled'' LDC run, and the bottom row are the results for ``independent'' LDCs. The solid horizontal lines on the Lomb-Scargle periodograms represent the false alarm probabilities of 10, 5, and 1~per~cent from bottom to top, calculated using {\sc astropy} routines \citep{astropy:2013, astropy:2018}. For the ``independent'' LDCs, the O-C plot has a reduced chi-squared value of $\chi_R=$0.87, whilst the ``coupled'' LDC O-C gives $\chi_R=$0.85. We find that there are no periodicities of statistical significance within the O-C data for either the ``coupled'' or ``independent'' LDC runs, and consequently conclude that there is no evidence of TTVs that would be indicative of a second planet in the WASP-126 system, in agreement with the findings of \citet{WASP-126c_no-evidence}.

The approach to fitting taken by \tranf{} differs to both \citet{WASP-126b_TTVs} and \citet{WASP-126c_no-evidence}, and consequently this result can be taken to be an independent verification of the findings of \citet{WASP-126c_no-evidence}. \citet{WASP-126b_TTVs} uses a simultaneous 2nd order detrending polynomial but this is  is not explicitly constructed to be depth-conserving. They do, however, use LDTk in their handling of LDCs, but not to directly inform the likelihood of trial parameters. Instead, \citet{WASP-126b_TTVs} uses LDTk to find the highest likelihood LDC values for the host parameters and \textit{TESS} filter and fixes the values here. We note however that the host parameters used by \citet{WASP-126b_TTVs} for this do not exactly match those of \citep{WASP-126b_discovery}, as indicated in Table \ref{tab:WASP-126b_results}, and it is unclear where the alternative value of host metallicity originates from. 

\citet{WASP-126c_no-evidence} fit physical parameters to the light curves after independently detrending the raw data. We note that they freely fit the LDCs without using the \citet{KippingLimbDarkening} parameterisation built into \tranf{}, and do not use any host information to inform the likelihoods. This is reflected in the larger LDC uncertainties. \citet{WASP-126c_no-evidence} compares their final LDC values to those predicted by bi-linearly interpolating the LDC tables provided by \citet{claret_bloeman_2011} for the Cousins \textit{R} and \textit{I} bands and the Sloan Digital Sky Survey \textit{z} band and then averaging the results to approximate the \textit{TESS} filter, which suggests values of $u_0=0.30$ and $u_1=0.28$. These predicted values are closer to the results of the \tranf{} ``independent'' LDC run, which most closely resembles the approach of \citet{WASP-126c_no-evidence},  but are in significant tension with the results from the ``coupled'' LDC run. 

We therefore conclude that, by taking a different approach to the investigations of both \citet{WASP-126b_TTVs} and \citet{WASP-126c_no-evidence}, we have been able to independently verify the result of \citet{WASP-126c_no-evidence} and find no statistically significant TTV signatures and, consequently, no evidence of a second planet in the WASP-126~b system.

\begin{table*}
    \caption{The planetary and orbital parameters of WASP-126~b derived using \tranf{} in both ``coupled'' and ``independent'' LDC fitting modes on observations from sectors 1--13, 27–-34, 36–-39, 61, and 63 of the \textit{TESS} mission. We present these alongside the values found by \citet{WASP-126b_TTVs}, \citet{WASP-126c_no-evidence}, and \citet{WASP-126b_discovery}. All fits assume zero orbital ellipticity, use a quadratic limb-darkening model, and unless otherwise stated assume the same host parameters as given by \citet{WASP-126b_discovery}. A total of 94,021 datapoints were fitted, and it gave a value of $f_e=1.17\times 10^{-04}$ and $\chi^2=89,956.47$ for the ``coupled" mode; and $f_e=1.21\times 10^{-04}$  and $\chi^2=89,898.40$ for the ``independent" mode.}
    \label{tab:WASP-126b_results}
    \centering
    \begin{tabular}{c c c c c c}
        \hline\hline
         & \tranf{}: coupled LDCs & \tranf{}: independent LDCs &\citet{WASP-126b_TTVs}& \citet{WASP-126c_no-evidence}& \citet{WASP-126b_discovery} \\
        \hline\\[-1em]
        $P$~[days] & $3.2887859^{+ 0.0000036}_{-0.0000035}~^\textrm{a}$ & $3.2887865^{+ 0.0000035}_{-0.0000039}~^\textrm{a}$ & $3.2888\pm1.94\times10^{-5}$& -~$^\textrm{c}$& $3.28880\pm0.00001$\\[.35em]
      
        $t_0$ [BJD$_{\textrm{TDB}}$]  & $2458327.52083^{+ 0.00058}_{-0.00068}~^\textrm{a}$ & $2458327.52070^{+ 0.00062}_{-0.00067}~^\textrm{a}$ &  - &$2456890.32004\pm0.00061$ & $2456890.3191 \pm 0.0006$~$^\textrm{d}$\\[.35em]
      
        $R_p/R_\star$ & $0.0776^{+ 0.0002}_{-0.0002}$ & $0.0778^{+ 0.0002}_{-0.0003}$ &  $0.0783\pm0.0002$& $0.07712^{+0.00063}_{-0.00047}$& $0.0781^{+0.0013}_{-0.0013}$\\[.35em]
        
        $R_p$ [$\rjup$] & $0.959^{+ 0.078}_{- 0.073}$ & $0.962^{+ 0.079}_{- 0.072}$ & $0.964\pm0.076$ & $0.953\pm0.075$ & $0.965\pm0.077$\\[.35em]
      
        $a/R_\star$& $7.771^{+ 0.099}_{-0.087}$ & $7.719^{+ 0.107}_{-0.082}$ &  $7.887\pm 0.040$ & $7.80^{+0.11}_{-0.20}$ & $7.63^{+0.64}_{-0.23}$\\[.35em]
      
        $a$ [AU] &$0.0459^{+0.0037}_{-0.0037}$ & $0.0456^{+0.0036}_{-0.0036}$ &  $0.0466 \pm 0.0037$ & $0.0461 \pm 0.0038$ & $0.0451\pm0.0052$\\[.35em]
    
        $i$ [deg] &$88.53^{+ 0.60}_{-0.39}$ & $88.27^{+ 0.57}_{-0.34}$ &   $89.51\pm 0.44$ & $88.7^{+0.9}_{-0.9}$ & $87.9^{+1.5}_{-1.5}$\\[.35em]
        
        $u_0$ &$0.366^{+ 0.017}_{-0.023}$ & $0.332^{+ 0.052}_{-0.028}$ &  $0.43~^\textrm{b}$ &$0.32^{+0.07}_{-0.07} $ & -\\[.35em]
        
        $u_1$ & $0.193^{+ 0.026}_{-0.026}$ & $0.229^{+ 0.069}_{-0.050}$ &  $0.14~^\textrm{b}$& $0.25^{+0.12}_{-0.13} $ & -\\[.35em]
        
        $T_{\text{eff},\star}$~[K] &- &- &- &- & $5800\pm100$\\
        
        $M_\star$~[$M_\odot$] &- & -& -&- & $1.12 \pm 0.06$\\
        
        $R_\star$~[$R_\odot$] & -& -& -& -& $1.27 \pm 0.08$\\
        
        ~[Fe/H] & -& -& $-0.06$ & - &$0.17 \pm 0.08$\\
        \hline
        \multicolumn{6}{l}{\footnotesize{$^\textrm{a}$ These values were derived assuming that no TTVs were present.}}\\
        \multicolumn{6}{l}{\footnotesize{$^\textrm{b}$ These values are those predicted by LDTk, using host parameters from \citet{WASP-126b_discovery} with $\textrm{[Fe/H]}=-0.06$.}}\\
        \multicolumn{6}{l}{\footnotesize{$^\textrm{c}$ The value of $P= 2.8493819\pm0.0000013$~days provided in \citet{WASP-126c_no-evidence} appears to be a typo as it exactly matches the period given for WASP-100~b.}}\\
        \multicolumn{6}{l}{\footnotesize{$^\textrm{d}$ The time standard is not specified.}}
    \end{tabular}\\
\end{table*}

\subsection{HST observations of WASP-43~b}
\label{sec:WASP-43b}
WASP-43~b, discovered by \citet{WASP-43b_discovery}, is a $2.034\pm0.052~\mjup$ hot Jupiter with a radius of $1.036\pm0.019~\rjup$, orbiting a type K7V star with a period of $0.813$~days \citep{WASP-43b_2012}. The proximity of the planet to the host star makes it an ideal candidate for emission spectroscopy, and it is one of the few exoplanets to have observed emission phase curve data \citep{WASP-43b_thermal_emission_HST}. Other studies have shown that WASP-43~b possesses a strong day-night temperature contrast \citep{WASP-43b_thermal_emission_HST, HyDRA_emission_spectra} and consequently strong equatorial jets in the atmosphere \citep{Kataria_equatorial_jets}. 

Since the first detection of an atmosphere by \citet{murgas_WASP-43b}, multiple transmission spectroscopy studies have been completed, with detections including water \citep{WASP-43b_kreidberg_HST}, carbon monoxide, carbon dioxide \citep{WASP-43b_CO2}, aluminium oxide \citep{WASP-43b_AlO}, and multiple hydrocarbon hazes \citep{WASP-43b_hazes}. Emission spectroscopy studies have also been utilised \citep{WASP-43b_thermal_emission_HST,WASP-43b_kreidberg_HST,WASP-43b_thermal_emission2014, stevenson2017spitzer}, and from these it has become apparent that a difference in the abundance of water exists between the day and night sides of the planet. 

Several of these studies, including those of \citet{Tsiaras_population_study} and \citet{WASP-43b_AlO}, make use of the \citet{WASP-43b_kreidberg_HST} observations from \emph{HST}, taken as part of observing program $13467$ \citep{WFC3_water_survey}. These observations were taken using the WFC3 G141 grism over the wavelength range of $1.135$--$1.642~\umu$m, and include three full-orbit phase curves, three primary transits, and two secondary eclipses. In this section, we demonstrate the application of \tranf{} to the observations in the $1.135$--$1.158~\umu$m waveband. We limit ourselves to this single waveband in this paper, as we are using \tranf{} to conduct an in depth study of observations of WASP-43~b from a wide range of sources (SPEARNET, in prep).

\begin{figure*}
    \centering
    \includegraphics[width=\textwidth]{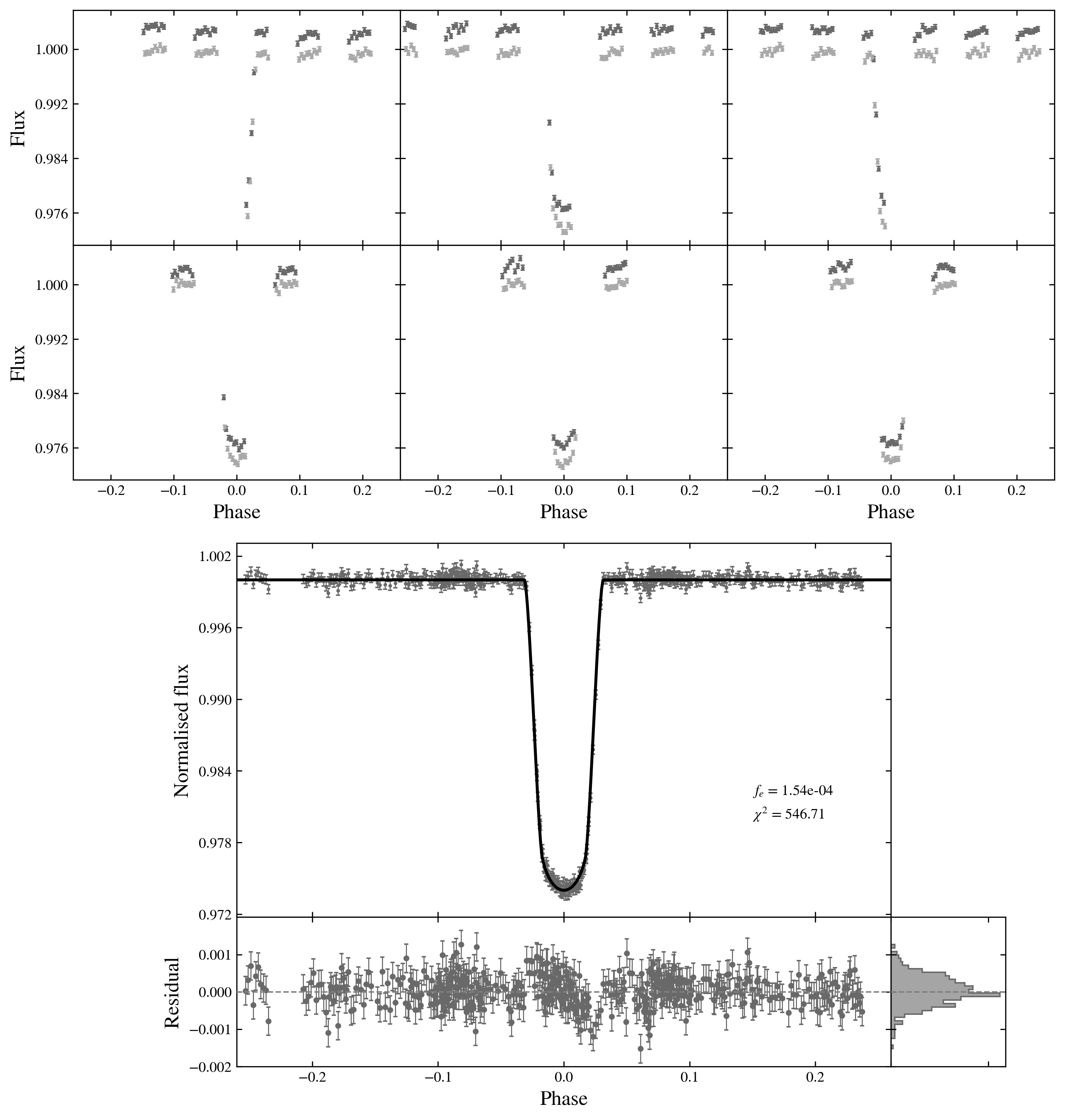}
    \caption{\emph{Top}: Observations of six transits of WASP-43~b taken using the Wide Field Camera 3 (WFC3) on \emph{HST} by \citet{WASP-43b_kreidberg_HST}. These data were all obtained using the $1.135$--$1.158~\umu$m waveband of the WFC3 G141 grism. The data from the first orbit of each visit and the first exposure of each orbit have been discarded and are not displayed or used in the subsequent analysis. The data are displayed here in raw form, normalised to a median value of $1$ and phase folded using the \tranf{}-retrieved ephemeris given in Table \ref{tab:WASP-43b_results_single_filter}. The scan direction used when obtaining the data introduces a systematic offset between the forward scan data (in dark grey) and the reverse scan data (in light grey). \emph{Bottom}: The phase-folded light fitted curve obtained from the above data using \tranf{} with a wavelength-coupled, quadratic limb darkening law and the custom detrending function given in Equation \ref{eq:HST_detrending}. The error bars have been scaled using the retrieved value of $f_e$. The best-fit light curve is over-plotted in black and the residuals are also displayed. We also report the $\chi^2$ corresponding to the fitting for a total of 469 datapoints.}
    \label{fig:WASP-43b}
\end{figure*}

The observations which include a transit are shown in the top plot in Figure \ref{fig:WASP-43b}, normalised to a median value of 1. For the purposes of this discussion, we distinguish between visits and orbits. A visit is a single one of these observations, and can be directly translated to an epoch within \tranf{}. Since \emph{HST} is an orbital satellite, there are times during a visit where the source cannot be seen, which leads to a temporal striping effect. Each one of these stripes is referred to as an orbit. We also define $t_{\rm visit}$, the time elapsed since the first exposure in a visit and $t_{\rm orb}$, the time elapsed since the first exposure in an orbit.

It can clearly be seen that, in addition to the observation-long trends, there are two complex systematics which affect the observations. These are the alternating offset introduced by the upstream/downstream effect \citep{HST_upstream_downstream}, and ramp-up effects in observations made in each orbit \citep{HST_ramp, charge_trapping}. Following \citet{Kreidberg_2018}, we assume that the observed flux, $F_{\rm obs}$ is given by 
\begin{equation}
    F_{\rm obs} = F_{\rm sig} \times F_{\rm sys},
\end{equation}
where $F_{\rm sig}$ is the astrophysical signal and $F_{\rm sys}$ is the signal from the systematics that need to be removed. We model these using 
\begin{equation}
    F_{\rm sys} = \left(S + v_1 t_{\rm visit} + v_2 t_{\rm visit}^2\right)\left( 1 - e^{-at_{\rm orb}-b} \right),
    \label{eq:HST_detrending}
\end{equation}
where
\begin{equation}
    S = 
    \begin{cases}
    1 & \textrm{ for forward scans}\\
    s & \textrm{ for reverse scans}    
    \end{cases}
\end{equation}
and $s$, $v_1$, $v_2$, $a$, and $b$ are all detrending coefficients. We note that \citet{Kreidberg_2018} multiply $S$ by a normalisation coefficient, but we exclude this as it would be degenerate with the normalisation constant fitted by \tranf{}. 

Implementing the detrending model in Equation \ref{eq:HST_detrending} as a custom detrending model within \tranf{}, we analyse the observations using the folded mode , ``coupled'' LDCs, and priors centred around \citep{WASP-43b_kreidberg_HST} results. The phase-folded lightcurve, residuals, and binned residuals are shown in the bottom plot in Figure \ref{fig:WASP-43b}, where it is clear that the systematics have been removed. In Table \ref{tab:WASP-43b_results_single_filter} we give the retrieved orbital parameters alongside those of \citet{WASP-43b_kreidberg_HST}. The temporal results are broadly consistent, but there is a discrepancy between the retrieved values of $a$ and $R_p$. We suggest that this could be caused by the slightly different approach used by \citet{WASP-43b_kreidberg_HST}. The \cite{WASP-43b_kreidberg_HST} detrending model is similar to our Equation (\ref{eq:HST_detrending}), but excludes the quadratic $v_2 t_{\rm visit}^2$ term from the visit-long trend, and it is not fitted simultaneously with their transit model. Additionally, there are only two free parameters used in the \cite{WASP-43b_kreidberg_HST} transit model; $R_p/R_\star$ and a linear LDC. Their period is fixed to that of \citet{WASP-43b_thermal_emission2014}, whilst their values for $i$, $a$, and $t_0$ are determined by fitting white-light observations. In our analysis all of the transit and de-trending parameters are fitted simultaneously by \tranf{}. Further investigation into these effects are being made as part of the aforementioned SPEARNET study, but we have included this section to demonstrate that \tranf{} is capable of fitting light curves that exhibit systematics as complex as those seen in observations made by \emph{HST}.

\begin{table}
    \centering
    \caption[Comparison of WASP-43~b parameters obtained by \tranf{} and \citet{WASP-43b_kreidberg_HST}]{The orbital and planetary parameters of WASP-43~b obtained by \tranf{} using the folded mode and wavelength-coupled quadratic LDC fitting on $1.135$--$1.158~\umu$m \emph{HST} observations, alongside the values used and obtained by \citet{WASP-43b_kreidberg_HST}. The only free parameters in the \citet{WASP-43b_kreidberg_HST} model are $R_p/R_\star$ and a linear LDC, the final value of which is not quoted.}
    \label{tab:WASP-43b_results_single_filter}
    \begin{tabular}{lcc}
        \hline\hline
                        & \textbf{\tranf{}}            & \citet{WASP-43b_kreidberg_HST} \\
        \hline\\[-1em]
        $P$~[days]      & $0.81348^{+0.00006}_{-0.00015}$ & $0.81347436$                            \\[.35em]
        $t_0$~[BJD$_{\textrm{TDB}}$]     & $2456601.0272^{+0.0020}_{-0.0005}$     & $2456601.02748$                         \\[.35em]
        $R_p/R_\star$   & $0.15764^{+0.00050}_{-0.00037}$            & $0.1597\pm0.0002$                       \\[.35em]
        $a/R_\star$     & $4.951^{+0.026}_{-0.034}$             & $4.872$                                 \\[.35em]
        $a$~[AU]        & $0.014^{+0.001}_{-0.001}$            & -                                       \\[.35em]
        $i$~[deg]       & $82.44^{+0.01}_{-0.76}$                & $82.1$                                  \\[.35em]
        $u_0$           & $0.242^{+0.025}_{-0.063}$              & -                                       \\[.35em]
        $u_1$           & $0.221^{+0.061}_{-0.086}$              & -                                       \\[.35em]
        \hline
    \end{tabular}
\end{table}

\subsection{Combined fitting of JWST and HST observations of WASP-96~b}
\label{sec:WASP-96b}
WASP-96~b is a hot Jupiter with a radius of $1.20\pm0.06~\rjup$ and mass of $0.48\pm0.03~\mjup$ orbiting a G8 type star with a period of $3.4$ days \citep{Hellier2014}. It has been a target for several atmospheric studies. An analysis of VLT FORS2 observations by \citet{Nikolov2018} showed an abundance of sodium in its atmosphere while predicting a cloud-free atmosphere at the limb. This analysis was reiterated in  \citet{Nikolov2022} while confirming oxygen signatures in its atmosphere using \emph{HST} WFC3 and \emph{Spitzer} IRAC data. Further, \citet{McGruder2022} also confirmed that the WASP-96~b atmosphere showed minimal aerosol content at its terminator. However, \citet{Samra2022} has predicted a presence of a cloudy atmosphere following an analysis of VLT, Spitzer, and \emph{HST} data. The atmosphere of WASP-96~b was shown to have features of water in its spectrum by \citet{Yip2020} using \emph{HST} observations, while also inferring a lack of high NH$_3$ in its atmosphere. As part of Early Release Observations of the \emph{JWST}, WASP-96~b was observed in NIRISS Single Object Slitless Spectroscopy (SOSS) mode consisting of 280 integration points, to demonstrate the capabilities of the instrument \citep{Pontoppidan2022}. In this section, we show the transmission spectrum of WASP-96~b generated from a \tranf{} analysis applied simultaneously to this \emph{JWST} data and to archive WFC3 \emph{HST} data.

\begin{figure*}
    \centering
    \includegraphics[width=\textwidth]{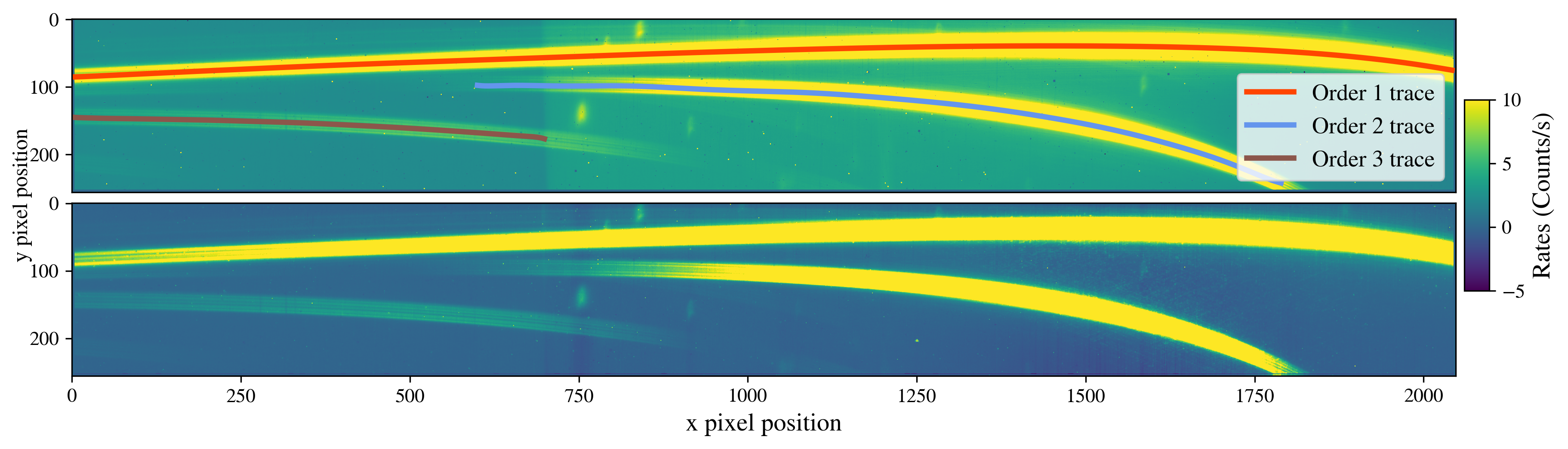}
    \caption{\emph{Top}: The traced spectra for Order 1 (red), 2 (blue), and 3 (brown) of the GR700XD grism are shown overplotted on the median image from WASP-96~b observations. \emph{Bottom}: The median of the image after noise correction.}
    \label{fig:WASP-96b_grism}
\end{figure*}

For the extraction of the lightcurves, the Stage 1 images from the NIRISS observations were processed following  \cite{Feinstein2023} and \emph{JWST} data analysis tools\footnote{\url{https://github.com/spacetelescope/jdat_notebooks/tree/main/notebooks}}. The images were corrected for bad pixels identified by DQ flags, by replacing them with the median value of the 9x9 pixel grid centred on them. After applying a flat-field correction, we traced the order 1, order 2, and order 3 spectra. The trace was then smoothed out by fitting it with a Chebyshev polynomial. The spectra of all three orders were masked out within 20 pixels from their traces and a noise and background correction was applied by removing the column median of the residual image. The traces of the spectra and the background-corrected versions are shown in Figure \ref{fig:WASP-96b_grism}. With an aperture radius of 15 pixels,
we extracted lightcurves from order 1 in 850-2800 nm range and from order 2 spectra in 600-850 nm range only, leaving out the noisier order 3 spectrum. The wavelength integrated lightcurve of this spectrum had a median noise of 720.1 ppm as compared to 77.0 ppm for order 1 and 164.0 ppm for order 2. The lightcurves were binned in wavelength bins of 25 nm, using inverse variance weighting.

\emph{HST} data from G141 and G102 grisms were extracted using Iraclis\footnote{\url{https://github.com/ucl-exoplanets/Iraclis}} \citep{tsiaras2016detection, iraclis}
in bins similar to that of \citet{Yip2020}. A total of 88 lightcurves obtained from \emph{JWST} observations and 38 lightcurves from \emph{HST} observations were fitted using \tranf{} to generate a transmission spectrum. We use a 2nd-order detrending function for \emph{JWST} lightcurves and a custom detrending function for \emph{HST} lightcurves as with the analysis of WASP-43~b described in Section~\ref{sec:WASP-43b}. The limb darkening coefficients were fitted in the coupled mode and using quadratic limb-darkening model. Planetary parameters and host parameters from \citet{Hellier2014} were used as priors and inputs, respectively, and are presented in Table \ref{tab:WASP-96b_results}. 
%For the wavelength-independent parameters, we report the average of the results using inverse variance weighting, from this run.
The ``batched'' mode was used for fitting these lightcurves, and the average of results weighted by inverse variance, were generated. Given the large number of lightcurves involved, there is a possibility of inter-batch variability in the wavelength-independent parameters. Consequently, the wavelength-dependent parameters might not result in the best-fit when used with the final results for wavelength-independent parameters. In order to reduce this discrepancy,  we run the ``batched'' mode again. To generate priors for this run, we take the results from the first run, and calculate inverse variance weighted results from all batches leaving one batch at a time. The union of these results is taken as the prior for the wavelength-independent parameters in the second run, which gives us the final results.

\begin{figure}
    \centering
    \includegraphics[width=\columnwidth]{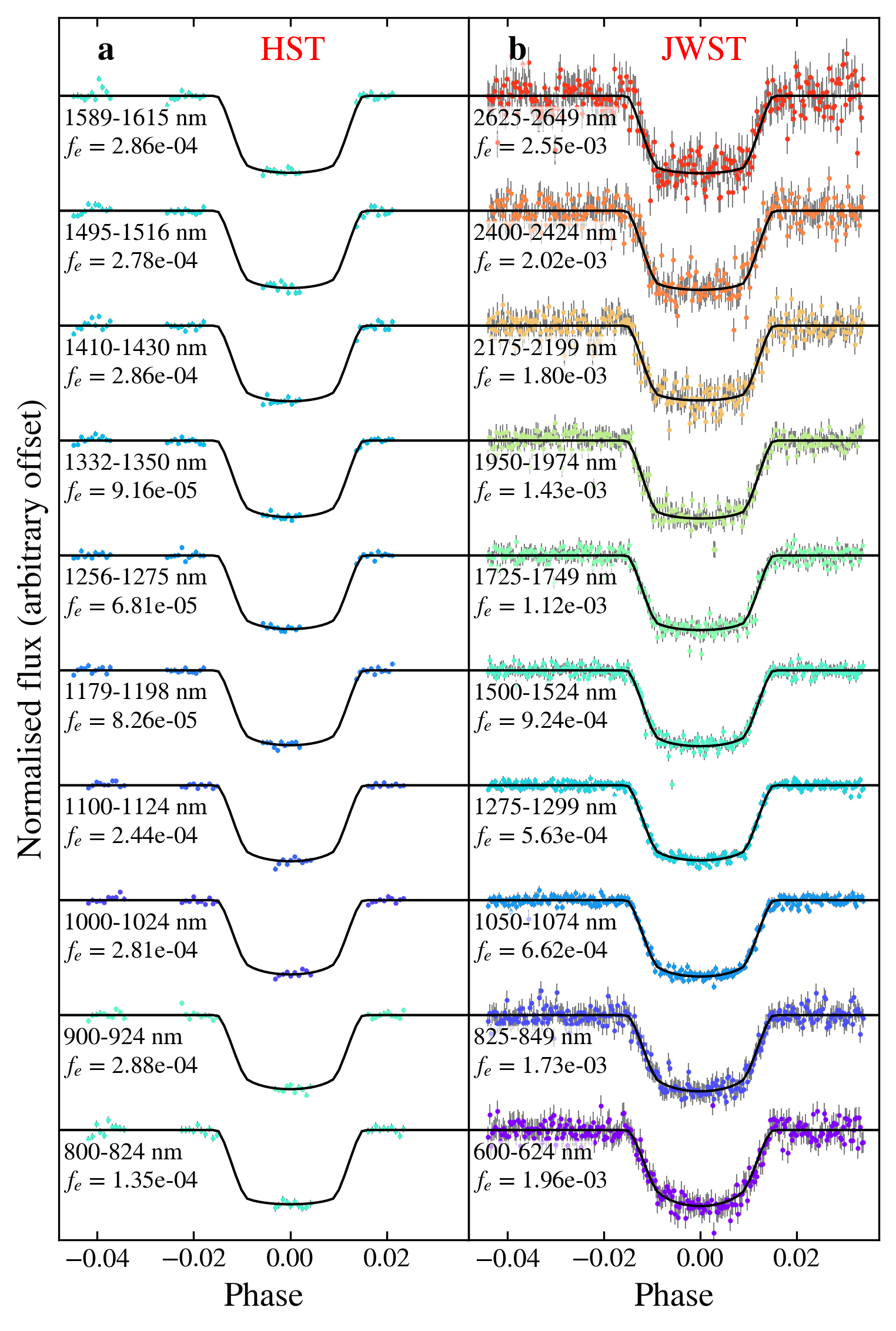}
    \caption{ Left column (a): a sample of fitted lightcurves from \emph{HST} observations of WASP-96~b, with the best-fit model overplotted in black. The error bars on all data points are plotted in grey and are scaled using the retrieved value of $f_e$. The corresponding wavelength bin for the lightcurves is indicated. The right column (b) shows the same for \emph{JWST} data. \tranf{} computed the best-fit model from both \emph{HST} and \emph{JWST} datasets simultaneously. The $\chi^2$ values for \textit{JWST} lightcurves were in the range of 269.38- 294.91 with an average of 280.08. For \textit{HST} lightcurves, the $\chi^2$ values were in the range of 38.07- 96.33 with an average of 53.89 for the data from G141 which had 44 data points. The data from G102 had 40 integration points and resulted $\chi^2$ values in the range of 38.46- 81.09 with an average of 59.22.} 
    \label{fig:WASP-96b_process}
\end{figure}

\begin{figure}
    \centering
    \includegraphics[width=\columnwidth]{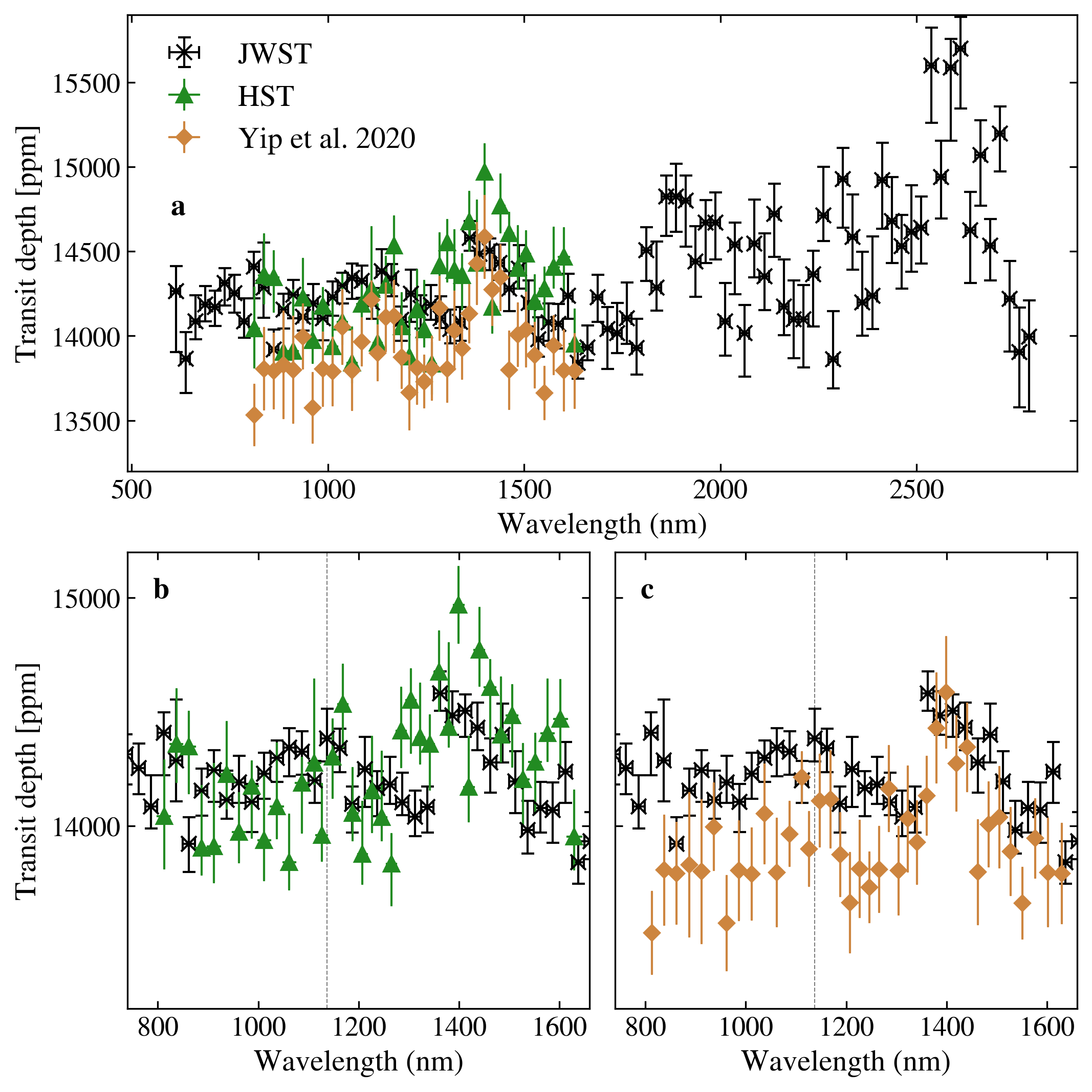}
    \caption{\textit{Top:} \tranf{} transmission spectrum of WASP-96~b derived simultaneously from \emph{JWST} (black) and \emph{HST} (green) observations. In orange, we also show for comparison the \emph{HST} transmission spectrum obtained by \citet{Yip2020}. \textit{Bottom:} zoomed panels spanning the \emph{HST} data range, showing the separate \tranf{} (\textit{left}) and \citet{Yip2020} (\textit{right}) solutions. The vertical dotted lines in these lower panels separate the two epochs of \emph{HST} observations.}
    \label{fig:WASP-96b}
\end{figure}

Figure \ref{fig:WASP-96b_process} shows the \emph{HST} and \emph{JWST} lightcurves after fitting \textbf{}, along with the best-fit model, with the best-fit values listed in Table \ref{tab:WASP-96b_results}, while the generated transmission spectrum is shown in Figure \ref{fig:WASP-96b}. The \tranf{} spectrum appears to be in good agreement with the analysis of \citet{Yip2020}.

\begin{table}
    \caption{The planetary and orbital parameters of WASP-96~b derived using \tranf{} analysis of the \emph{JWST} mission. The values are compared with the values retrieved from \citet{Hellier2014}, which were also used as priors for \tranf{}.}
    \label{tab:WASP-96b_results}
    \centering
    \begin{tabular}{c c c }
        \hline\hline
         & \tranf{} & \citet{Hellier2014} \\
        \hline\\[-1em]
        $P$~[days] & $3.4252519^{+0.0000011}_{-0.0000016}$ & $3.4252602\pm 2.7\times10^{-6}$\\[.35em]
      
        $t_0$ [BJD$_{\textrm{TDB}}$]  & $2458470.78034^{+0.00041}_{-0.00044}$ & $2456258.0621\pm 0.0002$~$^\textrm{a}$ \\[.35em]
      
        $a$ [AU] & $0.0441^{+0.0021
        }_{-0.0021}$ & $0.0453\pm0.0013$ \\[.35em]
    
        $i$ [deg] & $85.422^{+0.019}_{-0.018}$ & $85.6\pm 0.2$ \\[.35em]

        $R_p$ [$\mathrm{R}_{\mathrm{J}}$] & - & $1.20\pm 0.06$ \\
        
        $T_{\text{eff},\star}$~[K] &-& $5500\pm150$ \\
        
        $M_\star$~[$M_\odot$] &-& $1.06\pm0.09$ \\
        
        $R_\star$~[$R_\odot$] & -& $1.05\pm0.05$\\
        
        ~[Fe/H] & -& $0.14\pm0.19$\\
        \hline
        \multicolumn{3}{l}{\footnotesize{$^\textrm{a}$ This value is in UTC. The corresponding prior for $t_0$ was}}\\
        \multicolumn{3}{l}{\footnotesize{checked to be in-transit for the raw lightcurve.}}
    \end{tabular}\\
\end{table}

\section{Conclusions}
\label{sec:conclusions}
We have presented \tranf{}, a new open-source code for fitting exoplanetary transit light curves using nested sampling routines. \tranf{} has been designed for transmission spectroscopy surveys employing multiple telescopes, and allows coupling of limb-darkening coefficients across observation wavelengths by utilising information on the host star and the LDTk Python package \citep{Parviainen2015}. 

\tranf{} has been developed in anticipation of a new ``asset-starved'' era of transmission spectroscopy studies, where limited observational time and resources mean that studies will frequently have to combine data of various quality, wavelength coverage, and sources. One such example of this is SPEARNET, a survey which is using a heterogeneous distributed network of small- to mid-sized ground-based telescopes to conduct atmospheric studies of transiting exoplanets. 

Using \tranf{} and observations from the SPEARNET telescope network, we have presented analysis of new data of the hot-Neptune WASP-127~b, which includes the first $u'$-band observations of the planet. We have shown that introducing a wavelength-coupled approach to LDC fitting can result in changes as large as \textbf{8}~per~cent in the retrieved value of $R_p/R_\star$, or \textbf{17}~per~cent in measured transit depth. 

We have demonstrated the application of \tranf{} in more temporal-focused studies, analysing \textit{TESS} observations of 26 transits of WASP-91~b to produce updated planetary ephemerides.
This will prove invaluable in analysis of planets in the \textit{TESS} catalogue, allowing for easy searches for TTV signatures. We have used \tranf{} to analyse 180 transits of WASP-126~b observed by \textit{TESS} and have found no statistically significant evidence for the presence of TTVs proposed by \cite{WASP-126b_TTVs}. 

We have also shown how \tranf{} can be used in situations where observations display trends that cannot be adequately modelled with a low-order polynomial. We fitted observations of WASP-43~b from a single wavelength channel of \emph{HST} and found that the complex systematics can be adequately removed. Further analysis of the full set of \emph{HST} observations for WASP-43~b, in conjunction with observations from other telescopes, is being conducted as part of a separate study (SPEARNET, in prep). Moreover, we have already used \tranf{} to analyse the combined ground-based, \emph{HST} and \textit{TESS} observations of HAT-P-26~b \citep{2023arXiv230303610A}. The analysis found the presence of TTVs and H$_2$O dominated atmosphere on the planets, which agrees with previous studies. 

We used \tranf{} to construct a combined transmission spectrum of WASP-96~b from simultaneous fitting of \emph{JWST} NIRISS and \emph{HST} WFC3 data. In general we found good correspondence between the \emph{HST} and \emph{JWST} datasets. 

\section*{Acknowledgements}
    SPEARNET observations were made using ULTRASPEC at the Thai National Observatory and the Thai Robotic Telescopes, which is operated by the National Astronomical Research Institute of Thailand (NARIT, Public Organization).
    
    This paper includes data collected by the \textit{TESS} mission which is funded by the NASA Explorer Program, observations from NASA/ESA/CSA \emph{JWST} taken as part of proposal COM/ERO 2734 by the Early Release Observations Team, and data from \emph{HST} which is funded by NASA/ESA.

    Some of the structure of the \tranf{} code was inspired by the atmospheric retrieval code PLATON \citep{PLATON2019}.
    
    The authors would like to thank Nour Skaf for providing the spectral data for WASP-127~b in \cite{WASP-127b_2020b}, Guo Chen for providing the WASP-127~b data and model from \cite{WASP-127b_NaK&Li} and Laura Kreidberg for the \emph{HST} WFC3 data for WASP-43~b in \cite{WASP-43b_kreidberg_HST}.
    
    JJCH would like to thank the members of the JBCA Discord server, namely those in the \#help-me-senpai and \#tasty-science channels, for their help and insight in designing the plots for this paper.

    AP would like to thank Néstor Espinoza and Lili Alderson for suggestions on working with \emph{JWST} observations; and Andrew Mann for his insight on working with the errors on LDC.
    
    JJCH, JSM and AP are supported by PhD studentships from the United Kingdom's Science and Technology Facilities Council (STFC). EK and IM acknowledge support by the STFC grant  ST/P000649/1. VSD and ULTRASPEC are supported by STFC grant ST/V000853/1. This work is also supported by a National Astronomical Research Institute of Thailand (NARIT) research grant. TI and PM are supported by Development and Promotion of Science and Technology Talents Project (DPST), Thailand.

\section*{Data availability}
The codes underlying \tranf{} are available on GitHub at \url{https://github.com/SPEARNET/TransitFit}. The observational data of WASP-127~b obtained by SPEARNET and the \tranf{} transmission spectrum of WASP-96~b can be downloaded at \url{https://cdsarc.u-strasbg.fr/ftp/vizier.submit/transitfit_data/}. \emph{TESS}, \emph{JWST}, and \emph{HST} observations are publicly available from \url{https://mast.stsci.edu}. 



\bibliographystyle{mnras}
\bibliography{transitfit} % if your bibtex file is called example.bib
\appendix

\section{Posterior plots}
\label{sec:posterior_plot}
\begin{figure*}
    \centering
    \includegraphics[width=\textwidth]{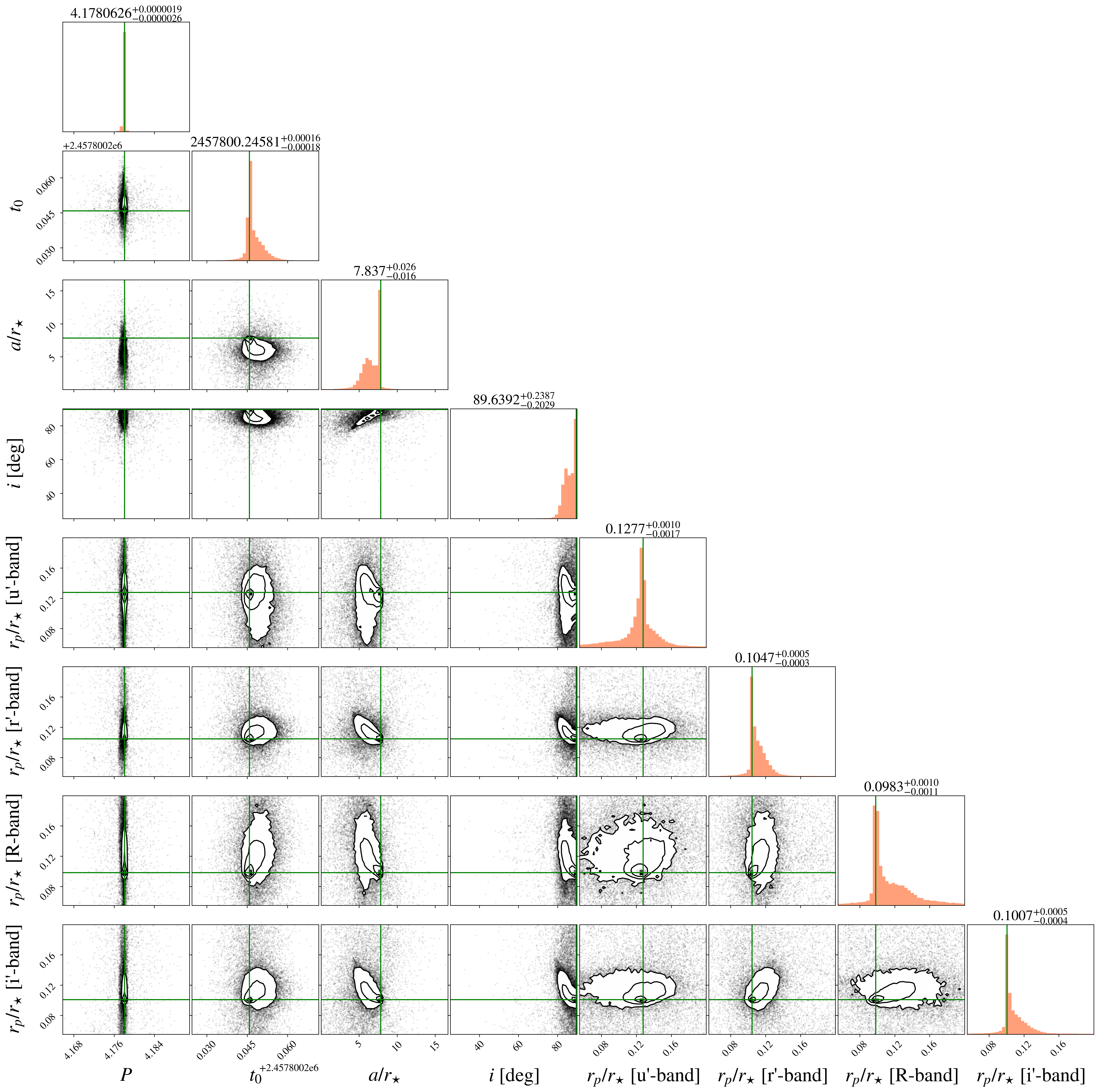}
    \caption{A posterior plot showing distribution of parameters samples across the range of priors. We use gaussian priors for $P$, $t_0$, and $a/r_{\star}$; and uniform priors for $R_p/R_{\star}$. The priors for inclination angle ($i$) are from a gaussian distribution clipped at a maximum of 90 degrees. These samples were taken for WASP-127~b fitting in `coupled' mode, and are plotted without weights. The LDCs, normalisation factor, and detrending coefficients are not shown for simplicity. The best fit values and corresponding error limits are shown as the titles of histograms.}
    \label{fig:posterior-plot}
\end{figure*}

\tranf{} returns the best-fit parameters which are the samples from {\sc dynesty} with highest likelihood. The fitting algorithm also generates posterior plots to show the density of parameters in the range of priors. A posterior plot is generated for each batch of fitting, and for the phase-pholded lightcurve if applicable. In Figure \ref{fig:posterior-plot} we have shown a posterior plot for selected parameters corresponding to WASP-127~b fitting in `coupled' mode.

%\begin{thebibliography}{99}
%\end{thebibliography}

% Don't change these lines
\bsp	% typesetting comment
\label{lastpage}
\end{document}